\pdfoutput=1
\documentclass[12pt]{emulateapj}
\usepackage{lscape}
\usepackage{subfigure}
\usepackage{graphicx}
\usepackage{multirow}
\usepackage{latexsym}
\usepackage{amssymb}
\usepackage{mathtools}
\usepackage{epsf}
\usepackage{threeparttable}

\newcommand{\kms}{km s$^{-1}$ }
\newcommand{\kmsn}{km s$^{-1}$}
\newcommand{\Msu}{$M_{\odot}$ }
\newcommand{\Msun}{$M_{\odot}$}
\newcommand{\degs}{$^{\circ}$ }
\newcommand{\degn}{$^{\circ}$}

\newcommand{\his}{{\rm H\,}{{\sc i }}}

\newcommand{\CO}{$^{13}$CO}
\newcommand{\COs}{$^{13}$CO }
\newcommand{\COJ}{$^{13}$CO J =  $ 1 \rightarrow 0 $ }
\newcommand{\COT}{$^{12}$CO}
\newcommand{\COTs}{$^{12}$CO }

\begin{document}	\title{Distribution and mass of diffuse and dense CO gas in the Milky Way}
%\author{\firstname{Julia}\surname{Roman-Duval}}
% \email{duval@stsci.edu}
% \altaffiliation{Space Telescope Science Institute} 
%\affiliation{3700 San Martin Drive, Baltimore, MD21218}
%\author{\firstname{James M.}\surname{Jacskon}}
% \email{jackson@bu.edu}
% \altaffiliation{institute for Astrophysics at Boston University} 
%\affiliation{725 Commonwealth avenue, CAS 512, Boston, MA 012215}

\author{Julia Roman-Duval\altaffilmark{1}, Mark Heyer\altaffilmark{2}, Chris Brunt\altaffilmark{3},  Paul Clark\altaffilmark{4}, Ralf Klessen\altaffilmark{5}, Rahul Shetty\altaffilmark{5}}
\altaffiltext{1}{Space Telescope Science Institute, 3700 San Martin Drive, Baltimore, MD 21218; duval@stsci.edu}
\altaffiltext{2}{Department of Astronomy, Lederle Research Building, University of Massachusetts, Amherst, MA 01003}
\altaffiltext{3}{Astrophysics Group, School of Physics, University of Exeter, Stocker Road, Exeter EX4 4QL, UK}
\altaffiltext{4}{School of Physics and Astronomy, QueenÕs Buildings, The Parade, Cardiff University, Cardiff, CF24 3AA, UK}
\altaffiltext{5}{Universit\"{a}t Heidelberg, Zentrum f\"{u}r Astronomie, Institut f\"{u}r Theoretische Astrophysik, Albert-Ueberle-Str. 2, 69120 Heidelberg, Germany}

\begin{abstract}
Emission from carbon monoxide (CO) is ubiquitously used as a tracer of dense star forming molecular clouds. There is, however, growing evidence that a significant fraction of CO emission originates from diffuse molecular gas. Quantifying the contribution of diffuse CO-emitting gas is vital for understanding the relation between molecular gas and star formation. We examine the Galactic distribution of two CO-emitting gas components, a high column density component detected in $^{13}$CO and $^{12}$CO, and a low column density component detected in $^{12}$CO, but not in $^{13}$CO. The ``diffuse'' and ``dense'' components are identified using a combination of smoothing, masking, and erosion/dilation procedures, making use of three large scale \COTs and \COs surveys of the inner and outer Milky Way. The diffuse component, which globally represents 25\% (1.5$\times10^8$ \Msun) of the total molecular gas mass (6.5$\times 10^8$ \Msun), is more extended perpendicular to the Galactic plane. The fraction of diffuse gas increases from $\sim$10-20\% at a galactocentric radius of 3---4 kpc to 50\% at 15 kpc, and increases with decreasing  surface density. In the inner Galaxy, a yet denser component traced by CS emission represents 14\% of the total molecular gas mass traced by \COTs emission. Only 14\% of the molecular gas mass traced by \COTs emission is identified as part of molecular clouds in \COs surveys by cloud identification algorithms. This study indicates that CO emission not only traces star forming clouds, but also a significant diffuse molecular ISM component.

\end{abstract}

\maketitle

\section{Introduction} \label{introduction}
\indent Stars are born from the fragmentation and collapse of dense cores within molecular clouds.  While the formation of stars within cores is dominated by gravity and is reasonably well understood, the mechanisms by which molecular clouds and molecular gas form and evolve remain an open question. For instance, it is not clear whether molecular clouds are long-lived gravitationally bound entities or transient over-densities in the underlying turbulent flow. The roles of radiative transfer, chemistry, magnetic fields and hydrodynamics  in shaping the structure and composition of molecular gas are also poorly constrained \citep{maclow04, Klessen14}. Understanding the physics of the molecular gas, and thereby the formation of stars, is crucial for comprehending galaxy formation and evolution.\\
\indent Molecular hydrogen (H$_2$) is an inefficient radiator within the cold environments of molecular clouds. Rotational emission from carbon monoxide (CO), the most abundant molecule in the dense phase after H$_2$, is widely used as a tracer of molecular gas instead. It is usually assumed that CO emission traces dense, well-shielded molecular gas that is or will be forming stars. However, there is growing evidence that a significant fraction of CO emission originates from relatively diffuse, non star-forming molecular gas. For example, \citet{goldsmith08} determine that 40\% of the molecular gas mass in the Taurus molecular cloud resides in diffuse molecular gas (N(H$_2$) $<$ 2.5$\times$10$^{21}$ cm$^{-2}$) that is detected in the \COTs line, but not the \COs line, and is not forming stars. Based on observations toward select sight-lines in the Milky Way, \citet{liszt10} determine a similar (40\%) fraction of diffuse non star-forming \COTs bright molecular gas. They conclude that the CO-to-H$_2$ conversion factor of this diffuse component is no different from the conversion factor of the dense gas ($X_{\mathrm{CO}}$ $=$ 2$\times 10^{20}$ cm$^{-2}$ K$^{-1}$ km$^{-1}$ s). In M51, \citet{pety13} quantify the distribution and mass of \COT-bright molecular gas, and conclude that 50\% of the CO emission originates from relatively low column density ($<$10$^{22}$ cm$^{-2}$) molecular gas on $\sim$kpc scales.\\
\indent  Studies of the Kennicutt-Schmidt (KS) relation \citep{schmidt59, kennicutt98} between molecular gas and star formation in Galactic \citep{heiderman10} and extragalactic \citep{krumholz12, shetty14a} environments also infer from the scale-dependence of the KS relation that a significant fraction of molecular gas must be in diffuse non star-forming phase. If the KS relation is sub-linear, the fraction of dense star-forming gas must decrease as the disk surface density increases, leading to longer molecular gas depletion times in higher surface density disks. Conversely, a super-linear KS relation implies that the dense gas fraction increases and that molecular depletion times decrease with increasing surface density. Previous studies of the KS relation in nearby galaxies have reported a range of KS slopes, from super-linear \citep{liu11, momose13}, to linear \citep{bigiel08, leroy13}, to sub-linear \citep{blanc09, ford13, shetty13, shetty14a}. Quantifying the contribution and distribution of diffuse CO-emitting molecular gas therefore  has important implications for our understanding of the processes leading to star-formation and thus to galaxy evolution. \\
\indent While the Milky Way offers the best spatial resolution to study this issue, quantifying the contribution of diffuse molecular gas is problematic in our own Galaxy. First, it is difficult to accurately estimate a distance to a parcel of molecular gas, due to the kinematic distance ambiguity, the large uncertainties on kinematic distances due to non-circular motions, and due to confusion in velocity space of near and far molecular clouds along the line-of-sight. To circumvent some of these issues, studies of the distribution and properties of molecular gas in the Milky Way \citep[e.g.,][]{solomon87, rathborne09, romanduval2010}  are forced to break up the CO emission in discrete molecular clouds identified by various available cloud identification algorithms such as CLUMPFIND \citep{williams94}, dendrograms \citep{rosolowsky08}, or GAUSSCLUMP \citep{stutzki14}. The drawback of this approach is of course that such detection algorithms would exclude diffuse CO emission. A detailed, high-resolution study of CO emission in the Milky Way is thus needed to better understand the spatial distribution of dense and diffuse molecular gas. This is possible with surveys acquired since the 2000s.  \\
\indent In this paper, we (re-)examine the luminosity and surface density distribution of CO-emitting gas in the inner (inside the solar circle) and outer (outside the solar circle) Milky Way, based on the Galactic Ring Survey (GRS) of \COs emission, the University of Massachusetts Stony Brook (UMSB) \COTs survey, and the Exeter-FCRAO (EXFC) survey (\COTs and \CO). In particular, we derive the spatial distribution (in luminosity and surface density) of three CO-emitting gas components in the Milky Way. Our study covers the Galactocentric radius range 3---15 kpc (and so excludes the Galactic Center). We identify gas that is detected in the \COTs line but shows no emission in the \COs line as the ``diffuse extended'' component. We define the ``dense'' component as the gas detected in both \COTs and \COs lines in the same voxel. Lastly, the ``very dense'' component corresponds to the gas detected in \COT, \CO, and carbon mono-sulfide (CS) 2-1 line emission. The CO-emitting gas components observed with these different tracers correspond to different density regimes, because their critical densities are different. The critical density of the \COTs and \COs 1-0 lines are similar at about $2\times10^3$ cm$^{-3}$, while the critical density of the CS 2-1 line is $5\times10^5$ cm$^{-3}$. However, due to optical depth effects (radiative trapping), the effective critical density of the \COT, \COs and CS lines are closer to $\sim$ $10^2$ cm$^{-3}$, $10^3$ cm$^{-3}$, and a few $10^4$ cm$^{-3}$. Additionally, we ensure that the S/N of the detection threshold is consistent for all 3 lines, so that the relative contributions of the three CO-gas components independent of the native sensitivities of the surveys. \\
\indent The paper is organized as follows. Section \ref{observations_section} describes the observations. In the subsequent section (\ref{method_section}), we describe the method to identify voxels (i.e., $\ell$, $b$, $v$ position) with significant emission, as well as estimate distances and other physical properties of the emitting regions, such as excitation temperatures and column densities. Section \ref{results_section} presents the derived properties, including the radial (with Galactocentric radius) and vertical (above and below the Galactic plane) distributions of diffuse, dense, and very dense components. Following a discussion of some limitations and implications of our analysis in Section \ref{discussion_section}, we conclude with a summary in Section \ref{conclusion_section}.

\begin{deluxetable*}{cccccc}
\tabletypesize{\scriptsize}
\tablecolumns{6}
\tablewidth{\textwidth}
\tablecaption{Parameters used for the categorization of voxels into ``noise'' and ``detection'', which includes smoothing, erosion, dilation, and thresholding}
\tablenum{1}
 
 \tablehead{
\multirow{2}{*}{} & \multicolumn{3}{c}{GRS+UMSB} & \multicolumn{2}{c}{EXFC 55-100 and 135-165}} \\

% & \colhead{\COT} &\colhead{\CO} &\colhead{CS}& \colhead{\COT} &\colhead{\CO} \\

 \startdata
 & \COT & \COs & CS & \COT & \CO \\
 &&&&&\\
\hline
&&&&&\\
 Original RMS per voxel & 0.47 K & 0.24 K & 0.26 K & 2.0 K & 0.70 K \\
Smoothing Kernel (voxels) & (1,1,3) & (3,3,7) &  (5,5,9) &(3,3,9) & (7,7,17)\\
Erosion dilation structure (voxels) & (5,5,7) & (5,5,7) & (5,5,7) & (5,5,9) & (7,7,17)\\
%Ratio of smoothed data sensitivities $\sigma(T_{MB}^{12, smoothed})/\sigma(T_{MB}^{13, smoothed})$ & \multicolumn{2}{c}{6.9}& &\multicolumn{2}{c}{5.8} &\\
Threshold & 1$\sigma$ & 1$\sigma$ &  1$\sigma$ & 1$\sigma$  & 1$\sigma$  \\

  \enddata
  \label{table_smoothing_parameters}
  \tablecomments{The size of a voxel is $1' \times 1' \times 0.3$\kms in the GRS+UMSB surveys, and 22.5''$\times$22.5''$\times$0.13 \kms in the EXFC survey}

%     \hline
\end{deluxetable*}

\begin{figure}
   \centering
               \includegraphics[width=8cm]{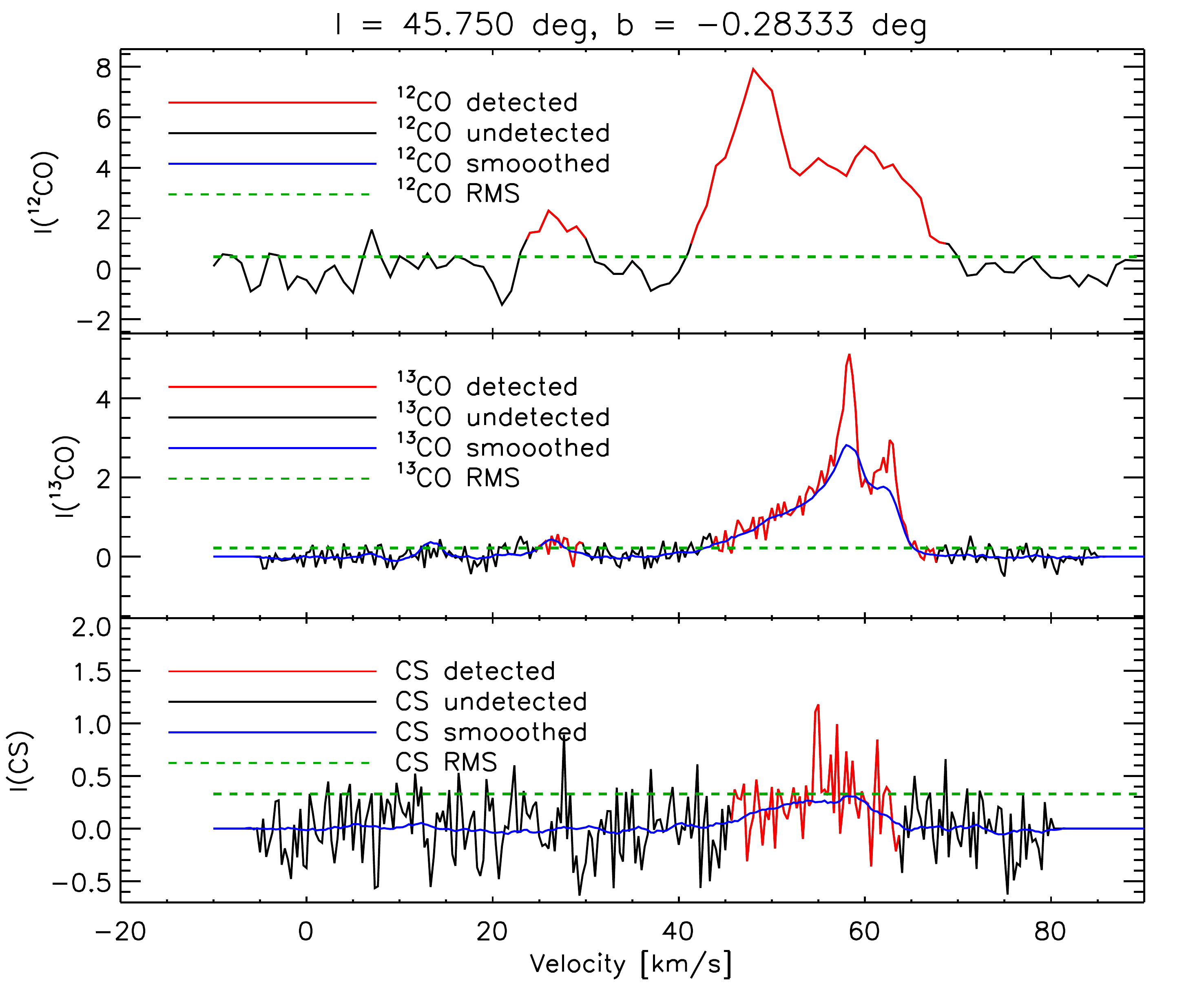} 
      \includegraphics[width=8cm]{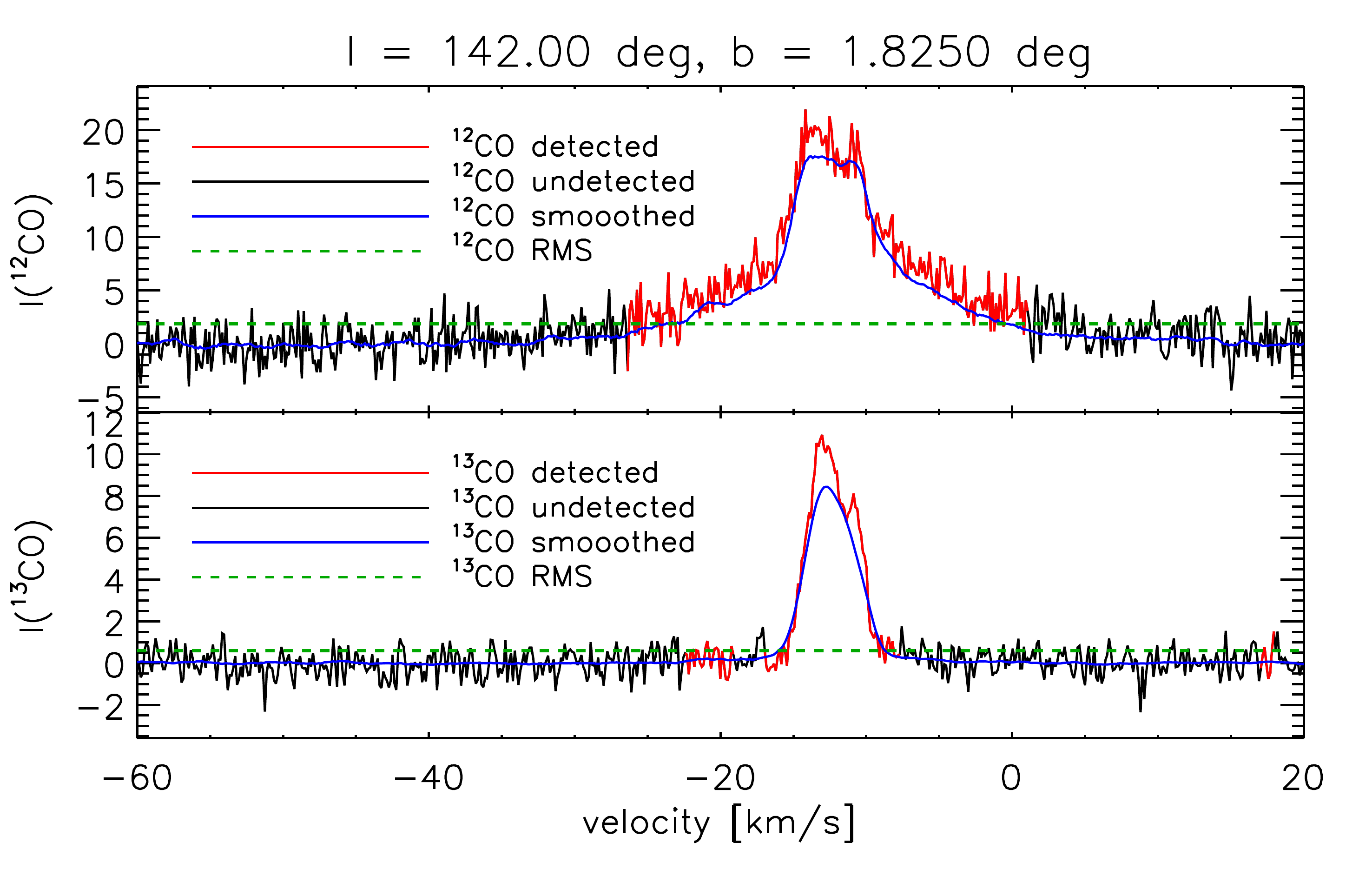} 
       \includegraphics[width=8cm]{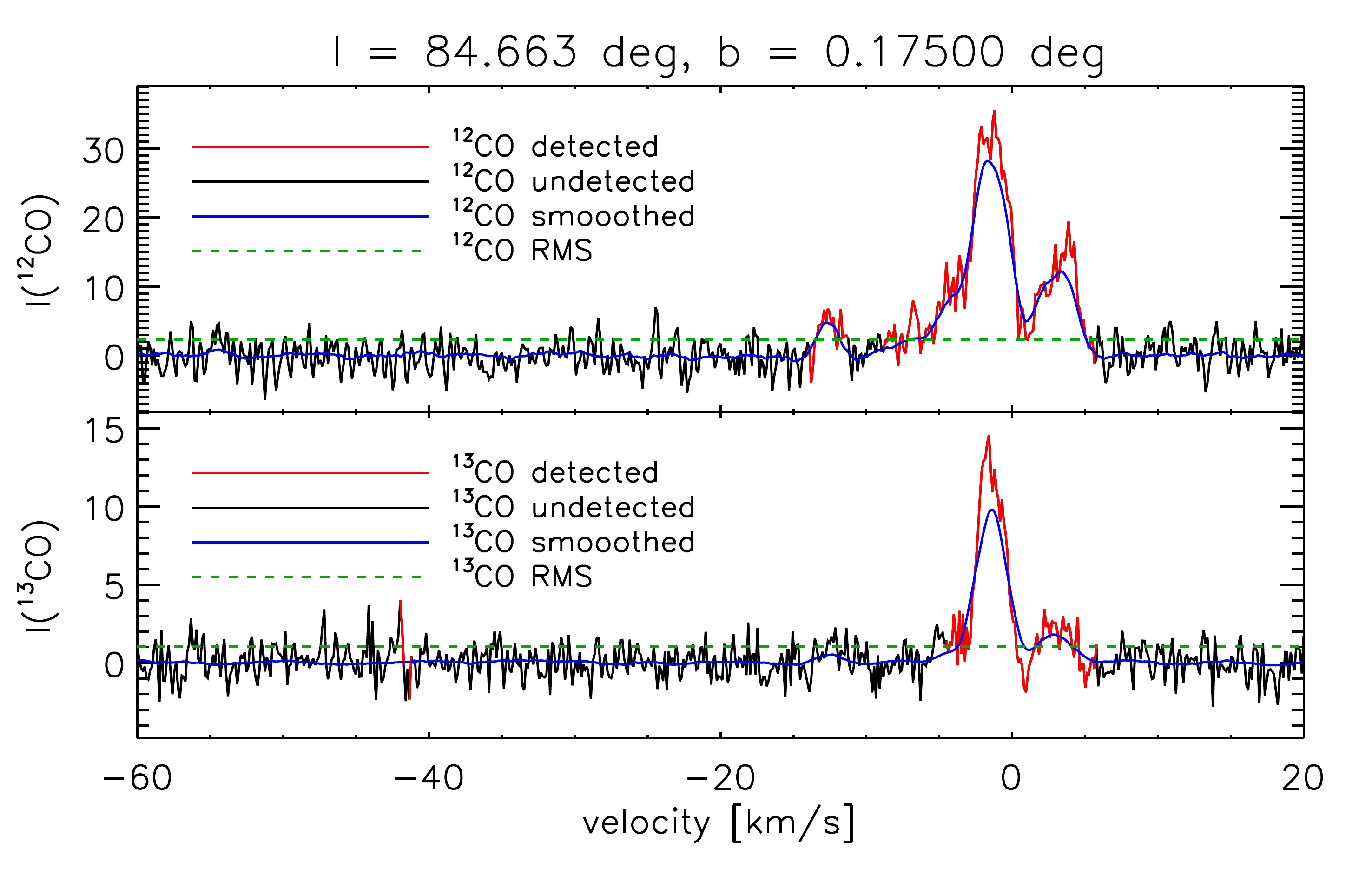} 

       \caption{Example of separation of voxels in the ``noise'' and ``detection'' categories in a sight-line of the GRS+UMSB surveys (top), in the EXFC 135-165 survey (middle), and in the EXFC 55-100 survey (bottom). The \COTs and \CO spectra are shown in the top two panels. For the GRS+UMSB only, the bottom panel shows the CS spectrum. The procedure described in Section \ref{det_noise_section} is used to compute the detection masks. The black and red lines indicate noise in the \COTs line and detected \COTs emission respectively. The blue curves correspond to the smoothed spectra. The dashed green line indicates the RMS main beam temperature of the un-smoothed spectra. }
\label{example_masking}
\end{figure}

\begin{figure*}
   \centering
               \includegraphics[width=8cm]{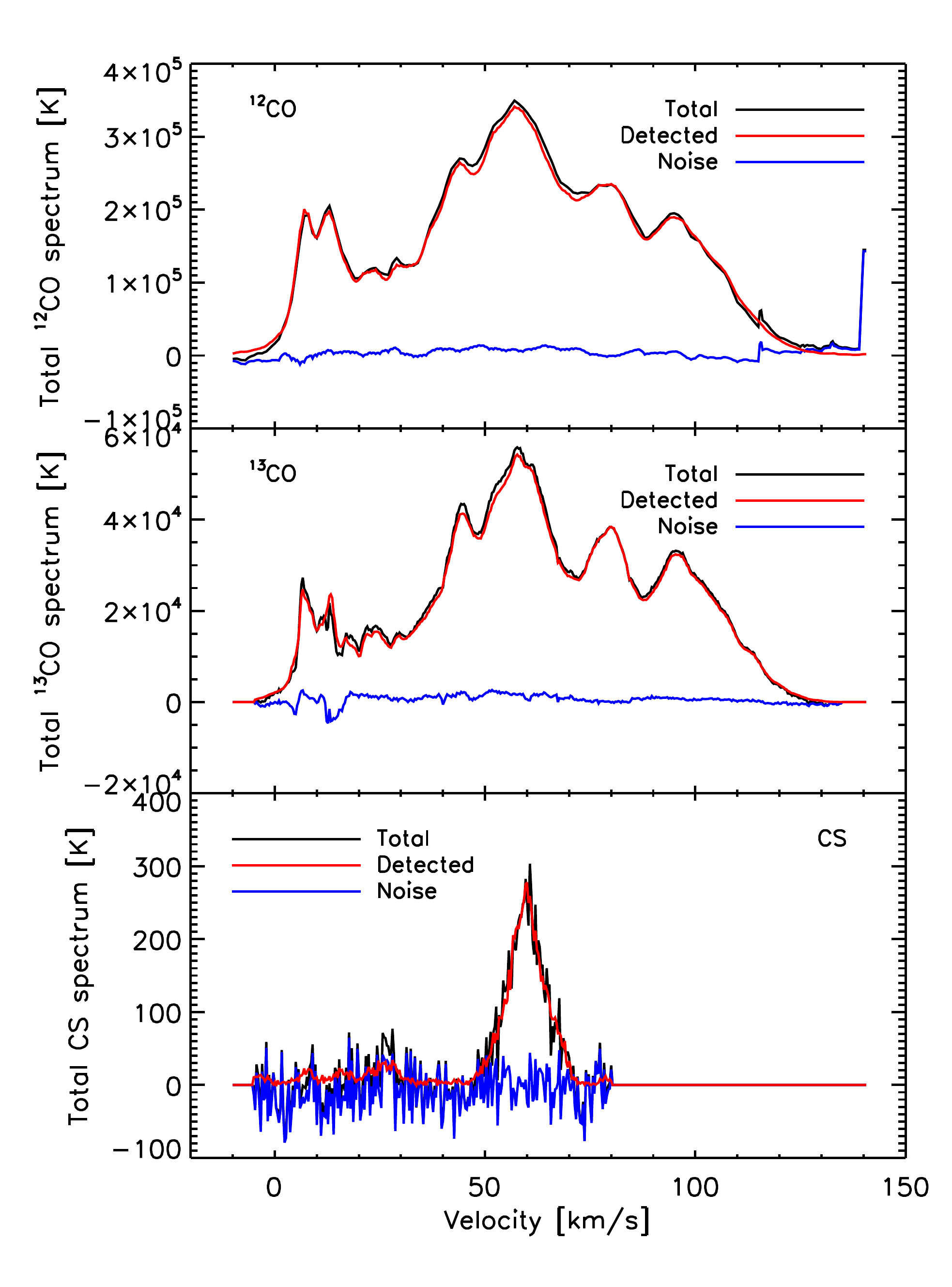} 
      \includegraphics[width=8cm]{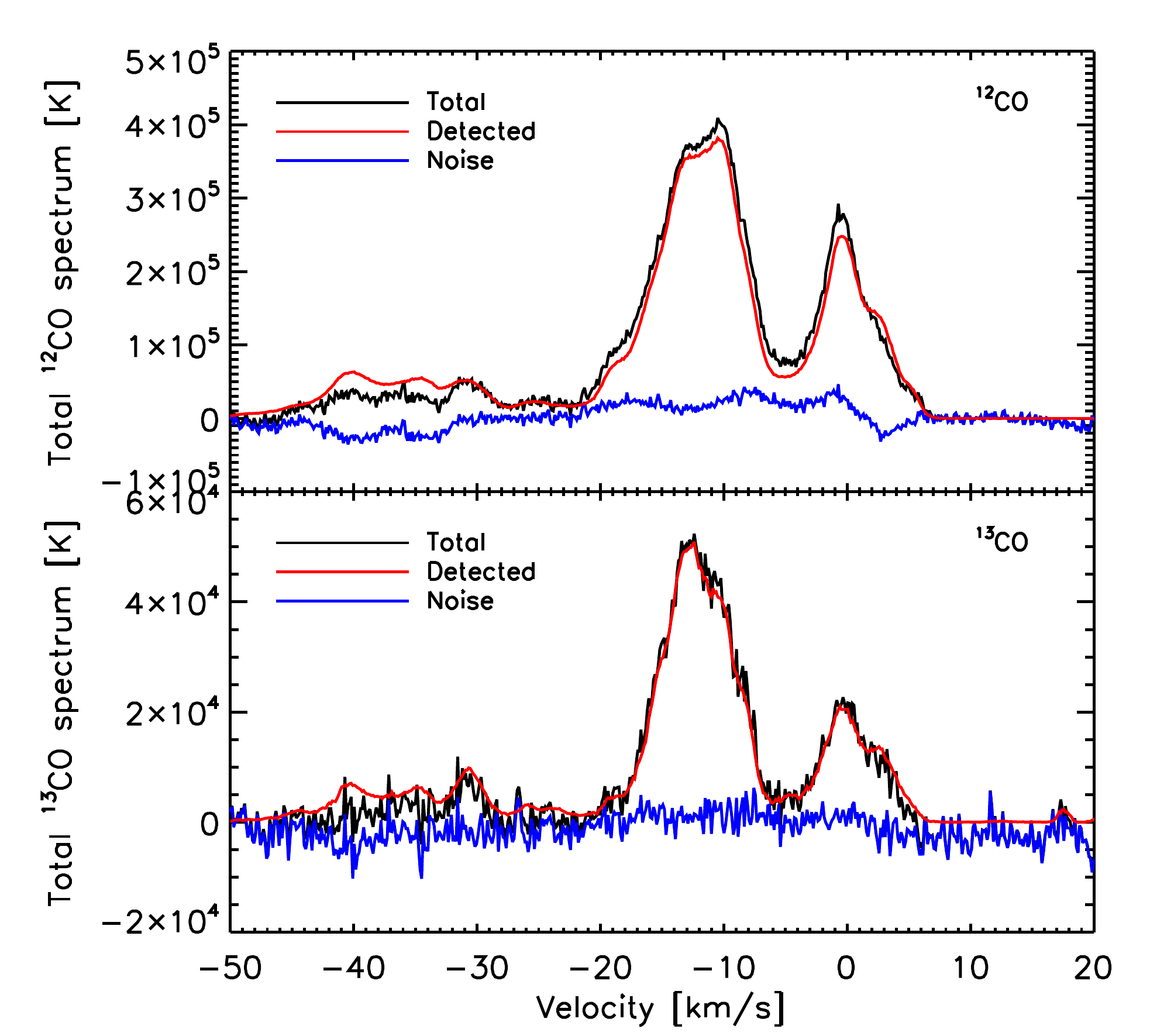} 
      \includegraphics[width=8cm]{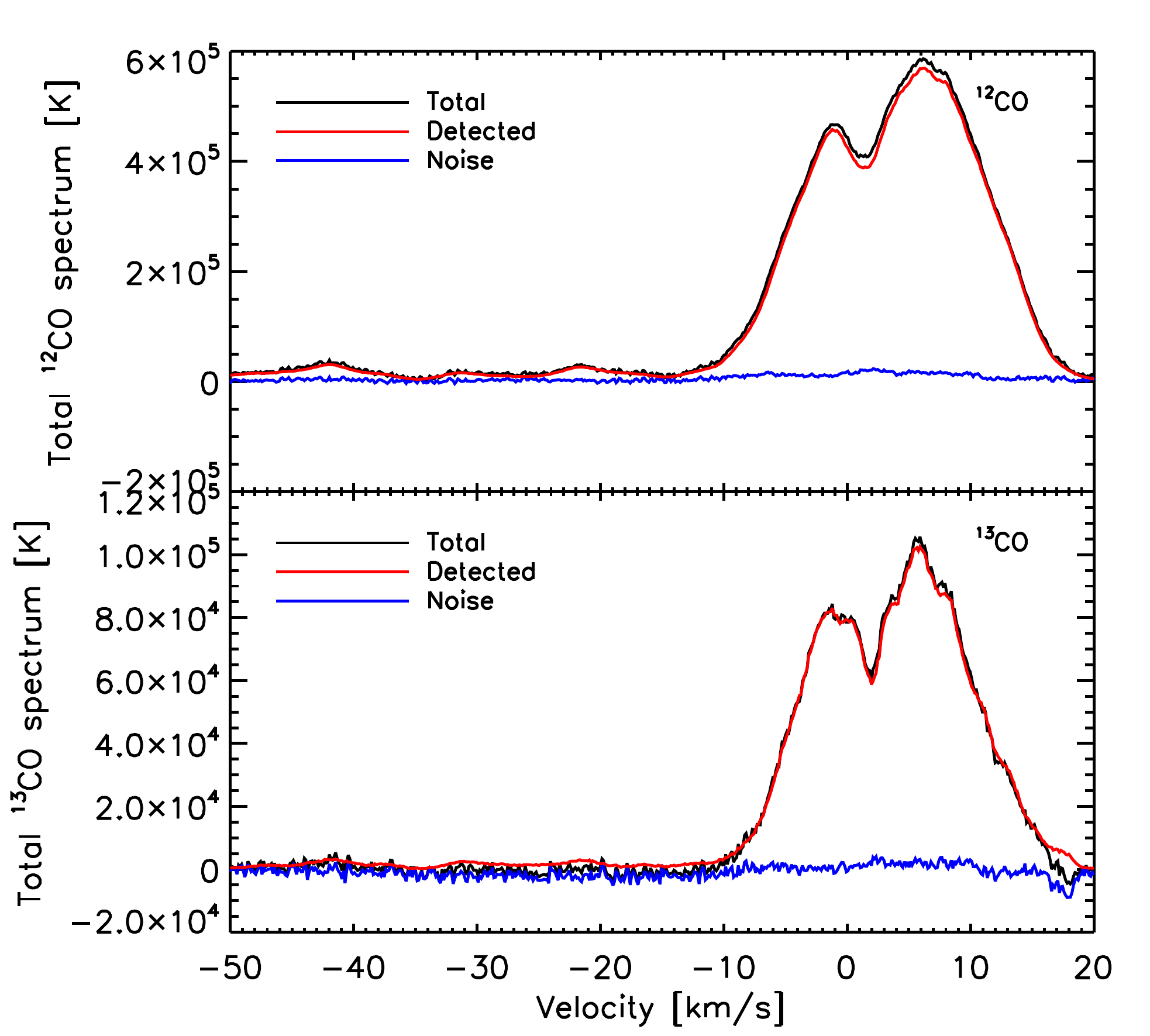}

       \caption{\COT, \CO, and CS spectra (inner Galaxy only) collapsed (summed) along the spatial dimensions, in the GRS+UMSB surveys (top left), in the $\ell$ $=$ 143\degs field of the EXFC 135-165 (top right), and in the $\ell$ $=$ 81\degs field of the EXFC 55-100 survey (bottom). The total spectra are shown in black. The collapsed spectra of ''detection'' voxels only are shown in red, and the total spectra of ''noise'' voxels are shown in blue.  The dip in the ''noise'' \COs spectrum in the GRS+UMSB at about 12-15 \kms is due to a problem with the off position in the GRS data, which causes an artificial absorption-looking feature in the baseline of the spectrum with main beam temperature values around $-$1.5---$-$1 K at longitudes $\ell$ $=$ 33---36\degn }
\label{collapsed_spectra}
\end{figure*}

\begin{figure*}
   \centering
               \includegraphics[width=\textwidth]{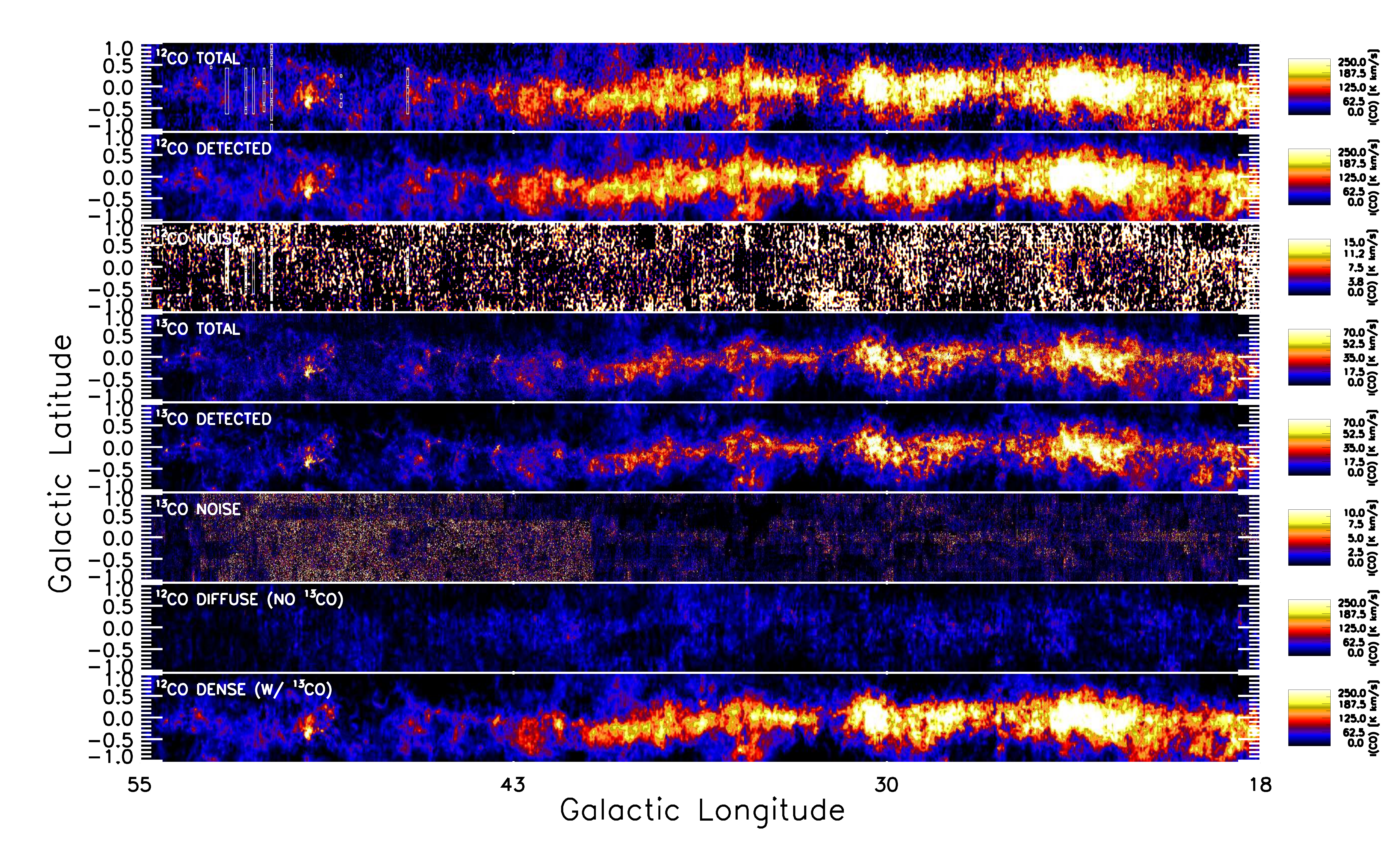} 
       \caption{From top to bottom, in the GRS+UMSB, integrated intensity maps of total \COTs emission (noise + detection), of the ''detected'' \COTs emission,  of the ''noise'' in the \COTs cube,  of the total \COs emission, of the detected \COs emission, of the noise in the \COs cube, and of the diffuse (\COT-bright and \CO-dark) and dense (\COT-bright and \CO-bright) \COTs components.}
\label{grs_maps}
\end{figure*}

\begin{figure}
   \centering
               \includegraphics[width=8cm]{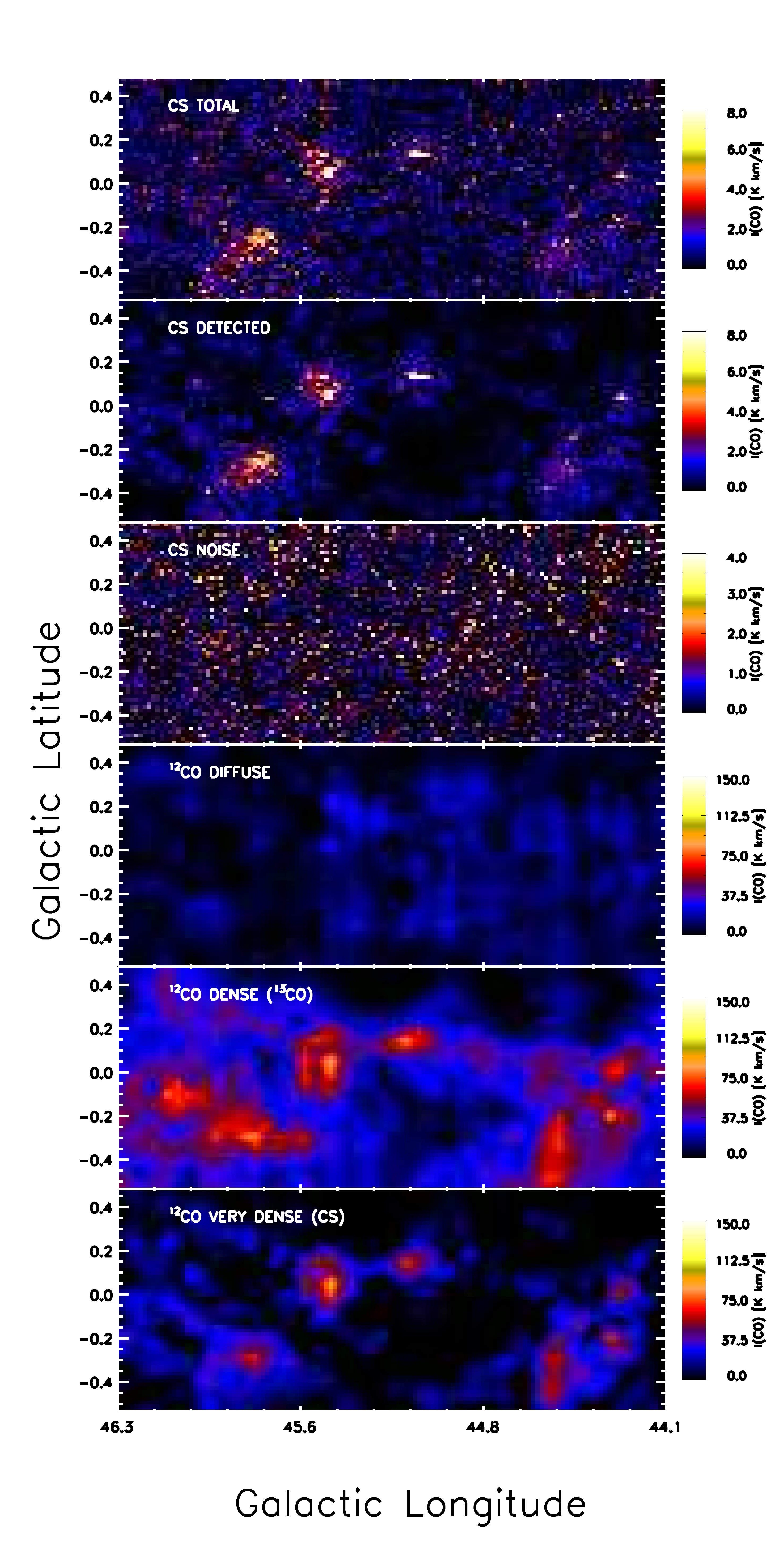} 
       \caption{From top to bottom, in the two square degree field of the GRS where CS is observed, integrated intensity maps of the total (noise+detection) CS emission, of the detected CS emission, of the noise in the CS cubes, of the diffuse (\COT-bright and \CO-dark), dense (\COT-bright, \CO-bright, and CS-dark), and very dense (\COT-bright, \CO-bright, and CS-bright)  \COTs components}
\label{cs_maps}
\end{figure}

\begin{figure*}
   \centering
               \includegraphics[width=\textwidth]{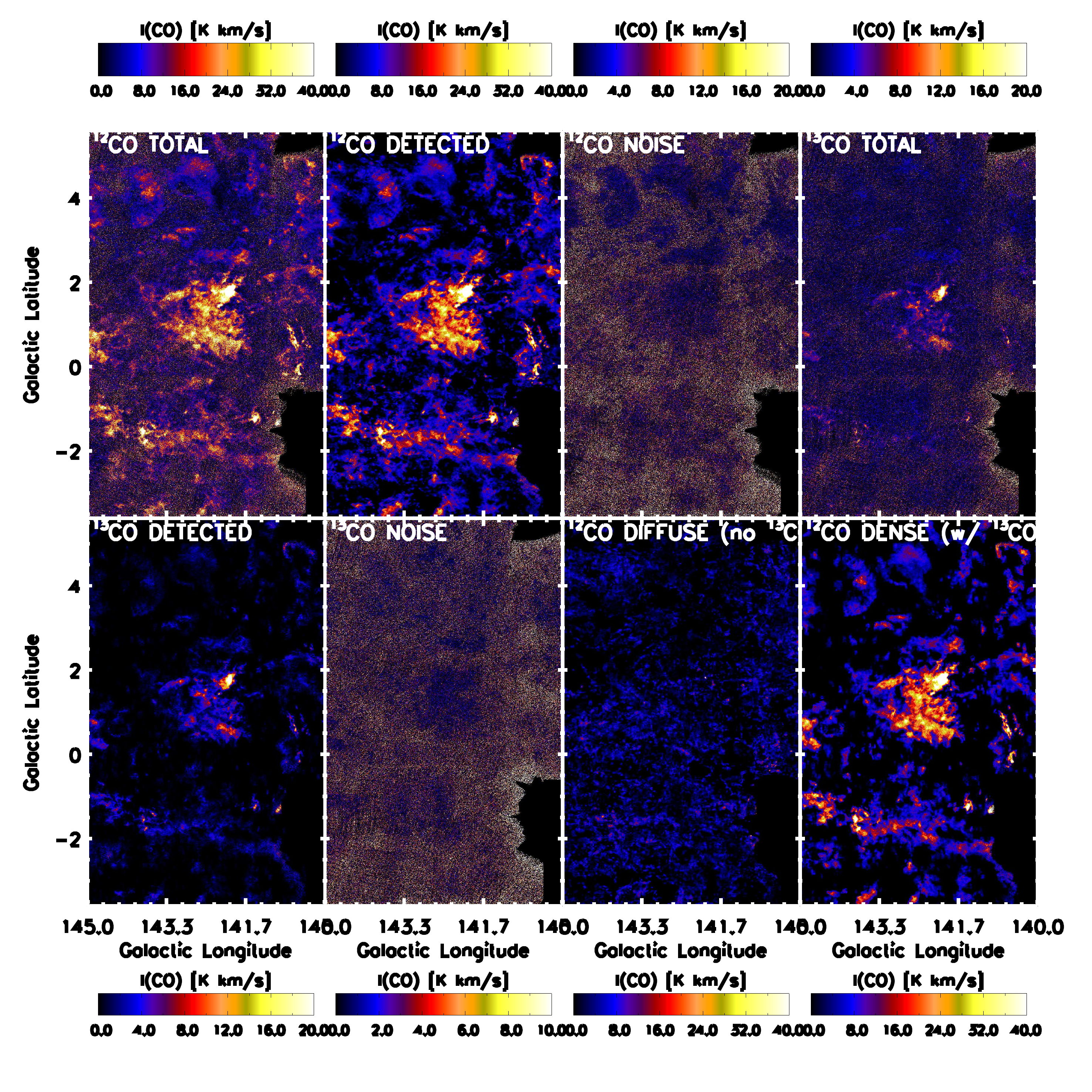} 
       \caption{Same as Figure \ref{grs_maps}, but for the $\ell$ $=$ 143\degs outer Galaxy field from the EXFC 135-165 survey}
\label{mark_maps1}
\end{figure*}

\begin{figure*}
   \centering
               \includegraphics[width=\textwidth]{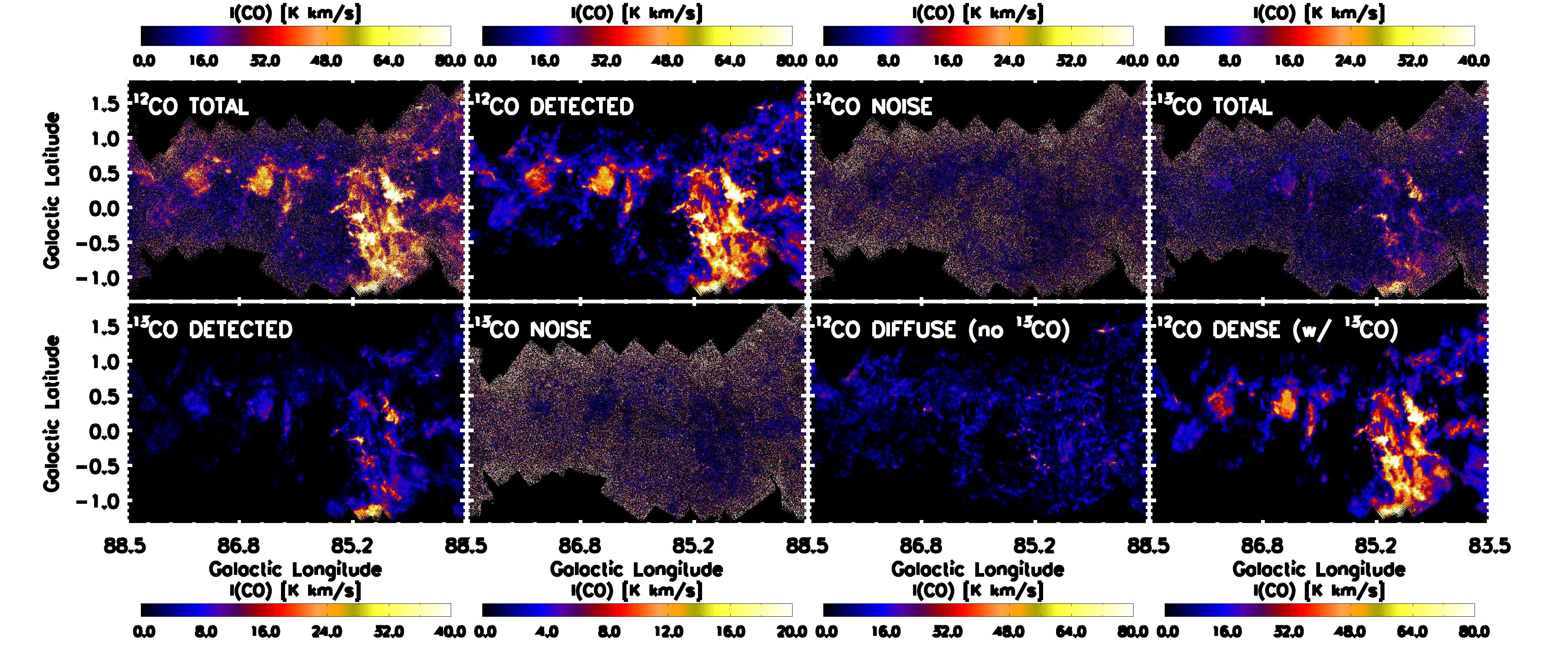} 
       \caption{Same as Figure \ref{grs_maps}, but for the $\ell$ $=$ 86\degs field from the EXFC 55-100 survey (inner and outer Galaxy)}
\label{mark_maps2}
\end{figure*}

\section{Observations} \label{observations_section}

\subsection{Observations of the $^{12}$CO J = 1 $\rightarrow$ 0 line in the inner Galaxy: University of Massachusetts Stony Brook Survey}
\indent In the inner Galaxy, the \COTs line was observed as part of the UMSB survey \citep{sanders86, clemens86}, a joint program between FCRAO and the State University of New York at Stony Brook performed between November 1981 and March 1984. All of the observations were obtained using the FCRAO 14 m telescope. A grid sampled every 3' covering the range 18\degs $<$ $\ell$ $<$ 55\degs and $-$1\degs $<$ b $<$ $+$1\degs  was observed with a velocity resolution of 1 \kms and an angular resolution of 45''. The UMSB survey covers the velocity range -10 \kms $<$ V$_{LSR}$ $<$ 140 \kmsn. The data were converted from  radiation temperature scale ($T_R^*$) to a main beam temperature scale ($T_{MB}$) via  $T_{MB}$ $=$ $T_R^*$/0.7.  \\

\subsection{Observations of the $^{13}$CO J = 1 $\rightarrow$ 0 line in the inner Galaxy: Galactic Ring Survey  (GRS)}
\indent The GRS survey observed a $\sim$ 40\degs section of the inner galaxy (18\degs $\leq$  $\ell$  $\leq$  55.7\degn, $-$1\degs $\leq$  $b$  $\leq$  1\degn ) in \COJ, using the FCRAO. The observations were taken between 1998 and 2005 with the SEQUOIA multipixel array. The survey achieved an angular esolution of 47'', sampled on a 22'' grid, and a spectral resolution of 0.212 \kms for a noise variance per voxel of $\sigma$(T$_A^*$) = 0.13 K ($\sigma_{T_{MB}} = 0.24$ K accounting for the main beam efficiency of 0.48). The survey covers the range of velocity $-$5 to 135 \kms for Galactic longitudes $\ell$ $\leq$ 40\degs and $-$5 to 85 \kms for Galactic longitudes $\ell$ $\geq$  40\degn. The data were converted from the antenna temperature scale $T_A^*$ to a main beam temperature scale $T_{MB}$ by correcting for the main beam efficiency of 0.48. \\

\subsection{Observations of the CS 2$\rightarrow$1 line in the inner Galaxy: Galactic Ring Survey  (GRS)}
\indent The GRS survey observed the CS 2$\rightarrow$1 line in 2 square degree field located at Galactic longitudes $\ell$ $=$ 44.3---46.3\degs and Galactic latitudes $b$ $=$ $-$0.5---0.5\degn, with the same velocity coverage as the \COs ($-$5 to 85 \kmsn). As for the \COs survey, the CS survey is also half-beam-sampled (45'' resolution with 22'' pixels). It achieved a sensitivity of $\sigma({T_{A}}^*) = 0.13$ K per voxel. The data were converted from the antenna temperature scale $T_A^*$ to a main beam temperature scale $T_{MB}$ by correcting for the main beam efficiency of 0.48. 

\subsection{Observations of the $^{12}$CO and $^{13}$CO J = 1 $\rightarrow$ 0 lines in the outer Galaxy: Exeter-FCRAO (EXFC) survey}
\indent Data for the EXFC survey (Brunt et al., in prep) were observed between 2003 and 2006 with the SEQUOIA beam array receiver. The survey spans two longitude ranges: $\ell$ $=$ 55-100\degs (hereafter EXFC 55-100), with the Galactic Latitude range $-$1.4\degs $\leq$ $b$ $\leq$ $+$1.9\degn, and $\ell$ $=$ 135-195\degs (hereafter EXFC 135-195), in the Galactic latitude range $-$3.6\degs $\leq$ $b$ $\leq$ $+$5.6\degn. The EXFC 135-195 survey covers the outer Galaxy only, while the EXFC 55-100 covers both the inner and outer Galaxy. \COTs and \COs 1-0 were observed simultaneously with angular resolutions of 45" and 48", sampled on a 22.5'' grid, and a spectral resolution of 0.127 \kmsn. The data were de-convolved to remove contributions by the antenna error beam and so are implicitly on a main beam temperature scale. We do not use the longitude range $\ell$ $=$ 165-195\degn, because the radial velocity of CO emission in this range is close to zero independent of distance (almost purely tangential motion).

\subsection{Mosaicking and regridding}
\indent The EXFC observations were split in 75 fields, sampling Galactic longitudes 55\degn---100\degs and 135\degn---195\degs every 3\degs. In this analysis, we do not use data with Galactic longitudes $\geq$ 165\degn, because at those longitudes, the motion of the gas is almost purely transverse (no radial velocity component), and a kinematic distance can therefore not be estimated robustly.  We re-sample and mosaic the full EXFC coverage below $\ell$ $=$ 165\degs into 13 disjoint mosaic fields spanning 5\degs in longitude and the full latitude range of EXFC ($-$3.6---5.6\degs for EXFC 135-195, $-$1.5---2.5\degs for EXFC 55-100). The individual spectra composing the mosaics are weighted by their RMS main beam temperature to produce the mosaics, and the original angular (22.5'') and spectral (0.127 \kmsn) sampling are conserved. Due to i/o and memory limitations, the entire survey cannot be stored into a single mosaic file.\\
\indent The GRS and UMSB surveys roughly cover the same area ($\ell$ $\simeq$ 18---55\degn, $b$ $\simeq$ $-$1---1\degn). However, small differences exist in the mapping strategy between the \COTs and \COs surveys, the GRS being half-beam-sampled on 22.5'' pixels, while the UMSB is under-sampled on a 3$'$ grid (both surveys being at 45'' resolution). To preserve some information about the spatial and spectral structure of the \COs and CS observations, we resample the GRS and UMSB on a common, intermediate grid with voxels 1$'\times 1' \times 0.3$ \kmsn. While interpolating the \COTs data does not improve its coarseness, it does allow us to more finely identify gas with \COTs emission and with (``dense'') or without (``diffuse'') \COs emission.

\subsection{Measurement uncertainties (noise RMS on main beam temperature)}
\indent An accurate estimation of the measurement errors is critical to this analysis. We compute a theoretical measurement error on the main beam temperature $T_{MB}$ (for \COT, \CO, and CS) at each $\ell$, $b$ position during the mosaicking process, based on the RMS of the data in its original form, and on the weights applied as part of the mosaicking process. However, residual (albeit small) baseline fluctuations between positions and within each spectrum can affect the noise RMS. We therefore empirically determine the noise RMS at each position on the sky using the following approach for the GRS and EXFC survey, in which the line emission is relatively sparse in the position-position-velocity (PPV) cubes. For each $\ell$, $b$ position, we determined the noise RMS (on the $T_{MB}$ scale) by fitting a Gaussian to the histogram of the spectrum (\COT,  \CO, or CS) in the range $min(T_{MB}) \leq T_{MB} \leq -min(T_{MB})$. The noise distribution is assumed to be Gaussian with zero mean, and therefore $min(T_{MB})<$ 0. The resulting standard deviation of the Gaussian provides an accurate value of the noise RMS of the observations. The fitted main beam temperature range ensures that most of the voxels included in the noise RMS measurement do not include actual CO or CS emission, which would bias the noise estimation. This procedure resulted in 2D maps of measurement errors for \COTs and \COs emission for EXFC, and \COs and CS emission for the GRS. Typical noise RMS values ($T_{MB}$ scale) in the GRS \COs, \COTs (EXFC), and \COs (EXFC) observations are 0.24 K per 1$'\times 1' \times$0.3 \kms voxel, 2 K per $22.5'' \times 22.5'' \times$0.13 \kms voxel, and 0.7 K per $22.5'' \times 22.5'' \times$0.13 \kms voxel, respectively.  \\
\indent In the inner Galaxy covered by the UMSB, which includes the molecular ring, the optically thick \COTs emission is ubiquitous and there are not enough voxels free of \COTs emission to estimate the RMS in each spectrum. Therefore, instead of using the histogram of individual sight-lines, we estimate the noise RMS from fitting a Gaussian to the histogram of the entire UMSB data set in the range $min(T_{MB}) \leq T_{MB} \leq |min(T_{MB})|$. The resulting noise RMS ($T_{MB}$ scale) for the UMSB is  $\sim$0.47 K per 1$'\times 1' \times$0.3 \kms voxel.

%\begin{figure*}
  % \centering
     % \includegraphics[width=\textwidth]{\dirgrs/display_RMS_grs_umsb_resamp3x} 
         %  \includegraphics[width=\textwidth]{grsu_rms} 	
       %\caption{RMS of the \COs main beam temperature in the GRS obtained by fitting a Gaussian to the distribution of main beam temperatures along each line-of-sight. The median RMS value is 0.24K per voxel. }
%\label{rms_maps}
%\end{figure*}

%\begin{figure}
   %\centering
      %         \includegraphics[width=8cm]{\dirgrs/plothist_umsb_noise} 

   %    \caption{Distribution of voxels with non-detections in the \COTs USMB survey, from which a constant baseline offset of 0.027K can be estimated and subtracted.}
%\label{umsb_histo}
%\end{figure}

\section{Method} \label{method_section}

\subsection{Distance calculation}
\indent Since mass and luminosity of CO-emitting gas are proportional to distance squared,  distances to each voxel (i.e., $\ell$, $b$, $v$ pixel location in the data) are required for our analysis. We compute kinematic distances to each voxel in the data, assuming that gas in the Galaxy rotates according to the  rotation curve derived by \citet{clemens85} for $R_{\odot}$ $=$ 8.5 kpc and $\theta_{\odot}$ $=$ 220 \kmsn. \\
\indent In the outer Galaxy, there is a single solution for the distance for a given radial velocity and Galactic longitude. The luminosity (in \COTs or \COs emission) of a voxel is therefore unambiguously determined. \\
\indent In the inner Galaxy ($R_{gal}$ $\leq$ 8.5 kpc, $v_{LSR}$ $\geq$ 0), there are two distance solutions for a given velocity, a ``near'' and a ``far'' distance. This is the well-know problem of the kinematic distance ambiguity.  The signal within a given voxel results from emission originating at either or both of those distances. Additional constraints are necessary to resolve this kinematic distance ambiguity. The \his self-absorption method \citep{knapp74, burton78, jackson02, romanduval2009} cannot be used for individual voxels, and we therefore take the following approach, which uses a Monte-Carlo simulation. For each of 10 statistically independent realizations, a near or far side distance is randomly assigned to each voxel based on the probability distribution of molecular gas height in the Galaxy.  Specifically, we assume that the vertical density profile within the molecular disk is a Gaussian with FWHM thickness of 110 pc \citep[see the review article by][references therein, and Section \ref{vertical_section}]{heyer15}.  The probability of molecular gas to be preset at height $z$ above or below the plane is also described by a Gaussian function with the same FWHM.  For each of the near and far distance solutions, we compute the height above the Galactic plane of a voxel given its Galactic latitude $b$, $z_{near, far} =  d_{near, far} \times tan(b)$, and the corresponding probabilities from the Gaussian vertical distribution $p_g(z_{near})$ and $p_g(z_{far})$. The relative probabilities of the emission in a voxel coming from the near and far distances are $p(near)$ $=$ $p_g(z_{near})$/$(p_g(z_{near}) + p_g(z_{far})$) and $p(far)$ $=$ $p_g(z_{far})$/$(p_g(z_{near}) + p_g(z_{far}))$. We then draw a random number from a uniform distribution between 0 and 1. If the random number is smaller than $p(near)$, the voxel is assigned to the near distance. Otherwise, it is assigned to the far distance. \\
\indent Once the distance is established, a CO luminosity and H$_2$ mass for each voxel is calculated. We save the 4-dimensional data cubes (mass, luminosity, distance, galactocentric radius) in ($\ell$, $b$, $v$, realization) space, and compute the spatial distribution of the luminosity and mass of CO gas for each Monte-Carlo realization. The spatial distribution of CO gas are then averaged between the different realizations to produce the figures in this paper, and the standard deviation between different realizations is included in the error budget. The standard deviation between realizations is very small compared to other sources of errors, and 10 realizations are more than what is necessary to obtain an accurate error estimation.\\
\indent  In reality, the signal in a voxel can originate from emission at both the near and far distances. The advantage of the Monte-Carlo method is that the final averaging between realizations distributes the signal in each voxel between the two distance solutions. \\
\indent For each voxel and each realization, an error on the distance (near or far) is also calculated. The error computation assumes 10 \kms non-circular motions, and computes the distance solutions $d_{near, far}^{\pm}$ for $v \pm 10$ \kmsn, where $v$ is the velocity of a voxel. The error on the distance is then $\delta d_{near,far}$ $=$ $|d_{near, far}^{+} - d_{near, far}^{-}|$/2. The distance error cubes ($\ell$, $b$, $v$, realization) are also stored and used in this analysis. The median errors on the near and far distances in the inner Galaxy are 25\% and 5\% respectively. In the outer Galaxy, the median error on the distance is 70\%. Generally, the fractional distance error increases with increasing longitude and with decreasing distance in the outer Galaxy.

\subsection{Identification of voxels with significant emission}\label{det_noise_section}

\begin{deluxetable*}{cccccc}
\tabletypesize{\scriptsize}
\tablecolumns{6}
\tablewidth{\textwidth}
\tablecaption{Number of voxels in the ``noise'' and ``detection'' categories for the \COT, \COs and CS lines in the GRS and UMSB ($\ell$ $=$ 18-55\degn)}
\tablenum{2}
 
 \tablehead{
& \COT & \COs & CS & Diffuse & Dense \-\
 }
 
 \startdata
Detection  & 3.1$\times 10^7$& 2.2$\times 10^7$ & 1.3$\times 10^5$ & 1.2$\times 10^7$ & 1.9$\times 10^7$\\
Noise & 9.6$\times 10^7$& 1.1$\times 10^8$ & 2.0$\times 10^6$ & --- & ---  \\
Filling factor (PPV space) & 24\% &17\% & 6\% &  39\% & 61\% \\
  \enddata
  \label{voxel_numbers_grs}

%     \hline
\end{deluxetable*}

\begin{deluxetable*}{ccccccccc}
\tabletypesize{\scriptsize}
\tablecolumns{9}
\tablewidth{\textwidth}
\tablecaption{Number of voxels in the ``noise'' and ``detection'' categories for the \COTs and \COs lines in the EXFC survey}
\tablenum{3}
 
 \tablehead{
 \multirow{2}{*}{} & \multicolumn{4}{c}{EXFC 135-195} & \multicolumn{4}{c}{EXFC 55-100} 
\\
 }
 
 \startdata
  %& \colhead{\COT} &\colhead{\CO} &\colhead{CS}& \colhead{diffuse} & \colhead{dense} &  \colhead{\COT} &\colhead{\CO} & \colhead{diffuse} & \colhead{dense} \\

&  \COT & \COs & Diffuse & Dense & \COT & \COs & Diffuse & Dense\\
&&&&&&&&\\
\hline
&&&&&&&&\\

Detection  & 1.0$\times 10^8$ & 5.0$\times 10^7$ & 5.1$\times 10^7$ & 5.0$\times 10^7$ & 1.1$\times 10^8$ & 7.4$\times 10^7$ & 3.9$\times 10^7$ &  7.4$\times 10^7$\\
Noise & 6.3$\times 10^9$ & 6.3$\times 10^9$ & --- & --- & 3.1$\times 10^9$ & 3.1$\times 10^9$ & --- & ---\\
Filling factor (PPV space) & 1.6\% & 0.8\% & 50\% &  50\% & 3\% & 2\% &  35\% & 65\% \\
  \enddata
  \label{voxel_numbers_exfc}

%     \hline
\end{deluxetable*}

\indent Our primary goal is to determine the spatial distribution, both in luminosity and mass, of CO-emitting  molecular gas. It is therefore crucial to capture the low-level extended emission. We are then faced with three difficulties. First, the sum of quantities (e.g., the main beam temperature or luminosity of a voxel) over a very large number of voxels (a single mosaic from EXFC contains approximately a billion voxels), which are potentially affected by small residual baseline offsets, can diverge or be dominated by those residual baseline effects. Second, capturing the low-level (low S/N) emission requires us to use a low threshold of detection (e.g., 1$\sigma$), which leads to positive biases in summed quantities, since positive noise peaks can be included and not their negative counterparts. Third, the relative contributions of those 3 components may depend on the signal-to-noise ratio (S/N) of the observations. For instance, if the S/N of the \COs observations is lower than that of the \COTs data, then the fraction of diffuse gas could potentially be inflated because of the inability to robustly detect \COs emission. \\
\indent To circumvent those difficulties, we have developed a robust method to categorize a voxel into ``noise'' or ``detection''. First, the spectral cubes are smoothed spatially and spectrally. The size of the smoothing kernels is determined so that the \COT, \CO, and CS data have similar S/N, which ensures that the relative fraction of the diffuse extended, dense, and very dense CO components relative to the total detected CO emission does not depend on the sensitivity of  the observations. We assumed the median RMS measurement error in each survey to compute the kernels sizes, and there is therefore one kernel size per survey and per line. Additionally, we assumed main beam temperature ratios $T_{12}/T_{13}$ $=$ 10 and $T_{12}/T_{CS}$ $=$ 15, based on the typical ratio observed in the line wings of individual spectra. These assumed ratios are only applied to determine the smoothing kernel widths and are not used for subsequent calculations of opacity. \\
\indent The $T_{12}/T_{13}$ ratio exactly defines the optical depth of the \COs line $\tau_{13}$ (see Section \ref{properties_subsection}), under the assumption that the beam filling factors of \COTs and \COs emission are the same, and that the excitation temperatures of the \COTs and \COs lines are also the same. Under these assumptions, our goal of detecting \COs in gas with $T_{12}/T_{13}$ $<$ 10 corresponds to $\tau_{13}$ $>$ 0.1. Variations in excitation temperature between the \COTs and \COs lines, and differences (possibly of a factor 2) in the beam filling factor of \COTs and \COs emission, could increase $\tau_{13}$ by several. \\
\indent Since there is a gradient in the \COT/\COs abundance with galactocentric radius \citep{milam05}, this target ratio $T_{12}/T_{13}$ corresponds to H$_2$ surface densities between 5 and 10 \Msu pc$^{-2}$ (km s$^{-1}$)$^{-1}$ s in the Galactocentric radius range probed here (3-15 kpc), at an excitation temperature of 8 K \citep[see][and Section \ref{properties_subsection}]{romanduval2010}.  Our goal of detecting CS emission with $T_{12}$/$T_{\mathrm{CS}}$ $<$ 15 corresponds to $\tau_{\mathrm{CS}}>0.07$. Assuming an abundance ratio $n(\mathrm{CS})/n(\mathrm{H}_2)$ $=$ 1$\times 10^{-9}$ \citep{neufeld15}, this implies the gas detected in CS emission has H$_2$ spectral surface densities $>$ 20 \Msu pc$^{-2}$ (km s$^{-1}$)$^{-1}$.\\
\indent Given these assumptions for the $T_{12}/T_{13}$ ratios, the sensitivity of the \COs smoothed cubes must be ten times better than the \COTs smoothed cubes in order for the fraction of diffuse and dense gas not to depend on the sensitivities of each spectral line data. Similarly, the CS smoothed cubes must be 15 times more sensitive than the \COTs smoothed observations. This constraint sets the number of elements in the smoothing kernels, via

\begin{equation}
\sqrt{\frac{N_{k13}}{N_{k12}}} = \left (\frac{T_{12}}{T_{13}} \right) \left (\frac{\sigma_{13}}{\sigma_{12}}  \right )
\end{equation}

\noindent where $N_{k12}$ and $N_{k13}$ are the number of voxels in the smoothing kernels for the \COTs and \COs cubes respectively, the line ratios are assumed as above, and ($\sigma_{13}/{\sigma_{12}}$) is the ratio of sensitivities in the un-smoothed \COTs and \COs cubes, taken to be the median RMS in each survey and line. A similar equation applies to the CS cubes. Once the number of elements in the kernels are determined, the elements must be distributed in the spatial and spectral directions. Several constraints determine the size of the kernel in each direction. First, the size of the kernels must be the same in the Galactic longitude and latitude directions. Second, the size of the kernels in each direction must be an odd number. Third, because the UMSB is spatially under-sampled, we must minimize the size of the smoothing kernels in the spatial direction. \\
\indent In the GRS+UMSB, where the native voxel is 1' $\times$ 1' $\times$ 0.3 \kmsn, we smooth the \COTs data with a (1, 1, 3) kernel, so $N_{k12}$ $=$3. The sensitivity of the smoothed \COTs cubes is 0.25 K per voxel (0.47/$\sqrt{3.}$). We choose a kernel of size (3, 3, 7) for the \CO, such that $N_{k13}$ $=$ 63, which allows us to probe the \COs line for $T_{12}/T_{13}$ ratios as high as 9 (close to the target value of 10). Similarly for the CS line, we can probe $T_{12}/T_{CS}$ $=$ 17 with a (5,5,9) kernel. In the EXFC survey, the native voxels are smaller (22.5'' $\times$ 22.5'' $\times$ 0.13 \kmsn), and the sensitivity per voxel is worse (2 K), and so the \COTs cubes are smoothed by a larger kernel of size (3, 3, 9) compared to the GRS+UMSB, corresponding to $N_{k12}$ $=$ 81. This ensures a sensitivity of 0.22 K per voxel in the smoothed \COTs cubes, consistent with the GRS+UMSB. We smooth the \COs cubes with a kernel of size (7, 7, 17) corresponding to $N_{k13}$ $=$ 225, and $T_{12}/T_{13}$ $=$ 9, also comparable to the GRS+UMSB. The sizes of the kernels are listed in Table 1.  \\
\indent In a second step, a detection mask is computed for each spectral cube (\COT, \CO, CS). The mask is equal to 1 where the smoothed spectral cube has a main beam temperature $T_{12}^{sm}$ (resp. $T_{13}^{sm}$) above 1$\sigma_{12}^{sm}$ (resp. $\sigma_{13}^{sm}$), where the noise RMS of the smoothed \COTs  (resp. \CO) cube $\sigma_{12}^{sm}$ (resp. $\sigma_{13}^{sm}$) is computed as the noise RMS of the original cube divided by the square root of the number of voxels in the smoothing kernel: $\sigma_{12}^{sm}$ $=$ $\sigma_{12}/\sqrt{N_{k12}}$ (and similarly for $\sigma_{13}^{sm}$). The mask is equal to zero everywhere else (non-detections).\\
\indent Because we only use a threshold of 1$\sigma$, a significant amount of spurious noise peaks are still included in the detection mask at this stage. This is problematic because only positive noise peaks are included in the masked data. When summing masses or luminosities over a large number of voxels, as we do here, these remaining noise peaks can significantly and positively bias the summed or binned quantities. To remove those noise peaks, the mask is eroded and then dilated by a structure of size similar to the smoothing kernels. This effectively removes sharp features (such as noise peaks) smaller than structure used in the erosion/dilation procedure. The ERODE and DILATE functions in IDL are used for this purpose. Erosion and dilation are morphological operations commonly used in image processing, and are described in, e.g., Soille (1999). Finally, the eroded/dilated mask is applied to the un-smoothed data to separate the cubes' voxels into ``detection'' and ``noise'' categories. The resulting number of voxels in each category (``noise'' or ``detection'') are listed in Tables 2 and 3 for the GRS+UMSB and EXFC surveys. \\
\indent Figure \ref{example_masking} shows examples of our detection procedure along one sight-line in each survey, with the total, detected, and noise spectra indicated by different colors. The velocity range in which the \COTs and \COs lines are detected extends to very low main beam temperature levels, and is similar between the two lines. This constitutes an additional verification that the \COTs and \COs lines are detected with similar S/N ratios. \\
\indent Figure \ref{collapsed_spectra} shows the total \COT, \CO, and CS spectra (summed along all sight-lines) in the ``detection'' and ``noise'' categories, as well as their total in each survey. For the EXFC survey, the $\ell$ $=$ 143\degs and $\ell$ $=$ 86\degs are shown. Since noise voxels dominate in number (see Tables 2 and 3), Figure \ref{collapsed_spectra} demonstrates that 1) there are no residual structures in the noise spectrum that resemble spectral lines, and our detection/masking algorithm has therefore successfully captured all the low-level extended emission, and 2) there is no thresholding-induced positive bias in the detected CO emission, which would appear as a systematically negative noise spectrum. We note that the dip in the ``noise'' \COs spectrum in the GRS at about 12-15 \kms is due to a contaminated ``off'' position in the GRS data, which causes an artificial negative feature in the baseline of the spectrum, with main beam temperature values around $-$1.5 to $-$1 at longitudes $\ell$ $=$ 33---36\degn. At some velocities, the ``detected'' spectrum is slightly larger than the ``total'' spectrum. This is due to small negative baseline fluctuations, and represents a very small effect, which is not seen in the combined fields. \\
\indent Figures \ref{grs_maps}, \ref{mark_maps1}, and \ref{mark_maps2} show integrated intensity images of the total, detected, and noise \COTs and \COs emission in each survey. The $\ell$ $=$ 143\degs and $\ell$ $=$ 86\degs fields are shown. Similarly, Figure \ref{cs_maps} shows the total, detected and noise CS emission in the two square degree field. Our algorithm produces much smoother and cleaner maps than if a naive sum along the velocity axis were performed. There is no residual structure in the noise maps, indicating that all the low-level emission was captured in the detection mask. \\

\subsection{Separation of the diffuse, dense, and very dense CO gas}

\indent Once ``noise'' and ``detection'' masks are created for the \COT, \CO, and CS cubes, we define the ``diffuse extended \COTs emission'' as the ensemble of all voxels where \COTs is detected, but \COs is not detected. The ``dense \COTs emission'' corresponds to all voxels where both the \COTs and \COs are detected. In the inner Galaxy field with CS observations, the ``very dense \COTs emission'' corresponds to voxels where \COT, \CO, and CS are detected. The definitions of the diffuse, dense, and very dense components are summarized in Table 4.  \\
 \indent The PPV cubes are smoothed to obtain the same S/N for the \COT, \CO, and CS observations. In Section \ref{det_noise_section}, we determined that the $T_{12}/T_{13}$ ratio of 10 assumed to derive the sizes of the smoothing kernels corresponds to H$_2$ surface densities of approximately 5-10 \Msu pc$^{-2}$ (km s$^{-1}$)$^{-1}$ (under certain assumptions, see Section \ref{det_noise_section}). Assuming a line width of 5 \kms typical of GMCs, this corresponds to surface densities of 25-50 \Msu pc$^{-2}$. Thus, by construction, we can detect the \COs line approximately down to ``spectral'' surface densities of 5-10 \Msu pc$^{-2}$ and the surface density threshold between ``diffuse'' and ``dense'' gas corresponds to H$_2$ surface densities of approximately 25-50 \Msu pc$^{-2}$. The density threshold between the diffuse and dense gas will vary depending on local conditions.\\
\indent Similarly, we can detect CS emission with $T_{12}/T_{\mathrm{CS}}<15$, which corresponds to H$_2$ spectral surface densities $>$ 20 \Msu pc$^{-2}$ (km s$^{-1}$)$^{-1}$, and surface densities $>$ 100 \Msu pc$^{-2}$. \\
\indent Thus, the ``diffuse'', ``dense'', and ``very dense'' components correspond to different surface density regimes. The approximate threshold surface densities of the ``diffuse'', ``dense'', and ``very dense'' gas are reported in Table 4.

\begin{deluxetable*}{ccccc}
\tabletypesize{\scriptsize}
\tablecolumns{5}
\tablecaption{Definition of the diffuse, dense, and very dense components}
\tablenum{4}
 
 \tablehead{
% & Diffuse & Dense & Very dense
& \COTs 1-0 & \COs 1-0 & CS 2-1 & $\Sigma_v$(H$_2$)\\
  & & & & (\Msu pc$^{-2}$ (km s$^{-1}$)$^{-1}$)\\
 }
 
 \startdata
Diffuse &  Detected & Undetected & Undetected & $<$ 10\\
Dense & Detected & Detected & Undetected & $>$ 10\\
Very dense & Detected & Detected & Detected  & $>$ 20\\

    \enddata
    \tablecomments{$\Sigma_v$(H$_2$) is the approximate threshold spectral surface density between the different regimes ``diffuse'', ``dense'', and ``very dense''.}
  \label{table_definitions}

%     \hline
\end{deluxetable*}

\subsection{Physical properties of each voxel}\label{properties_subsection}

\begin{figure}
   \centering
            \includegraphics[width=8cm]{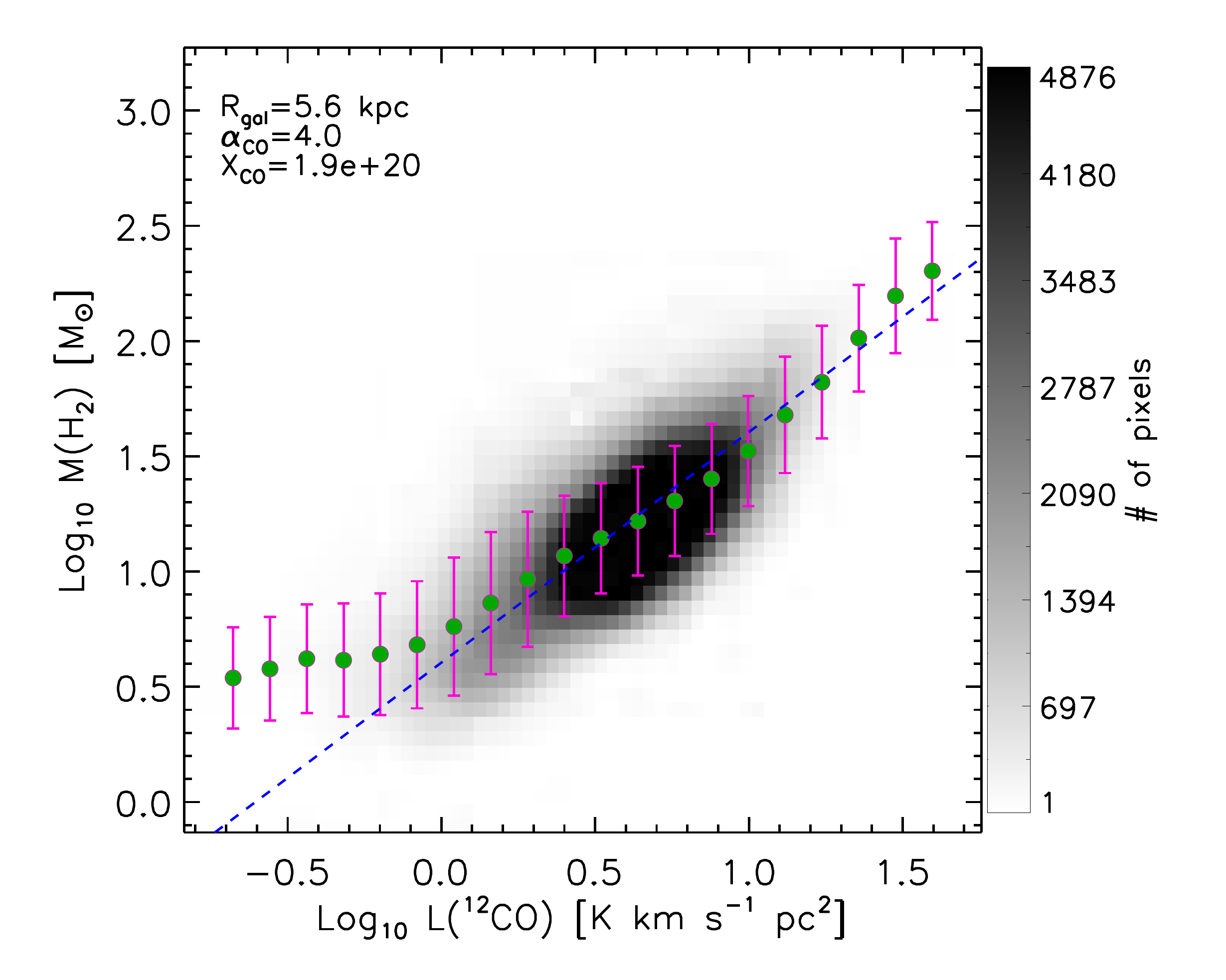} 
             \includegraphics[width=8cm]{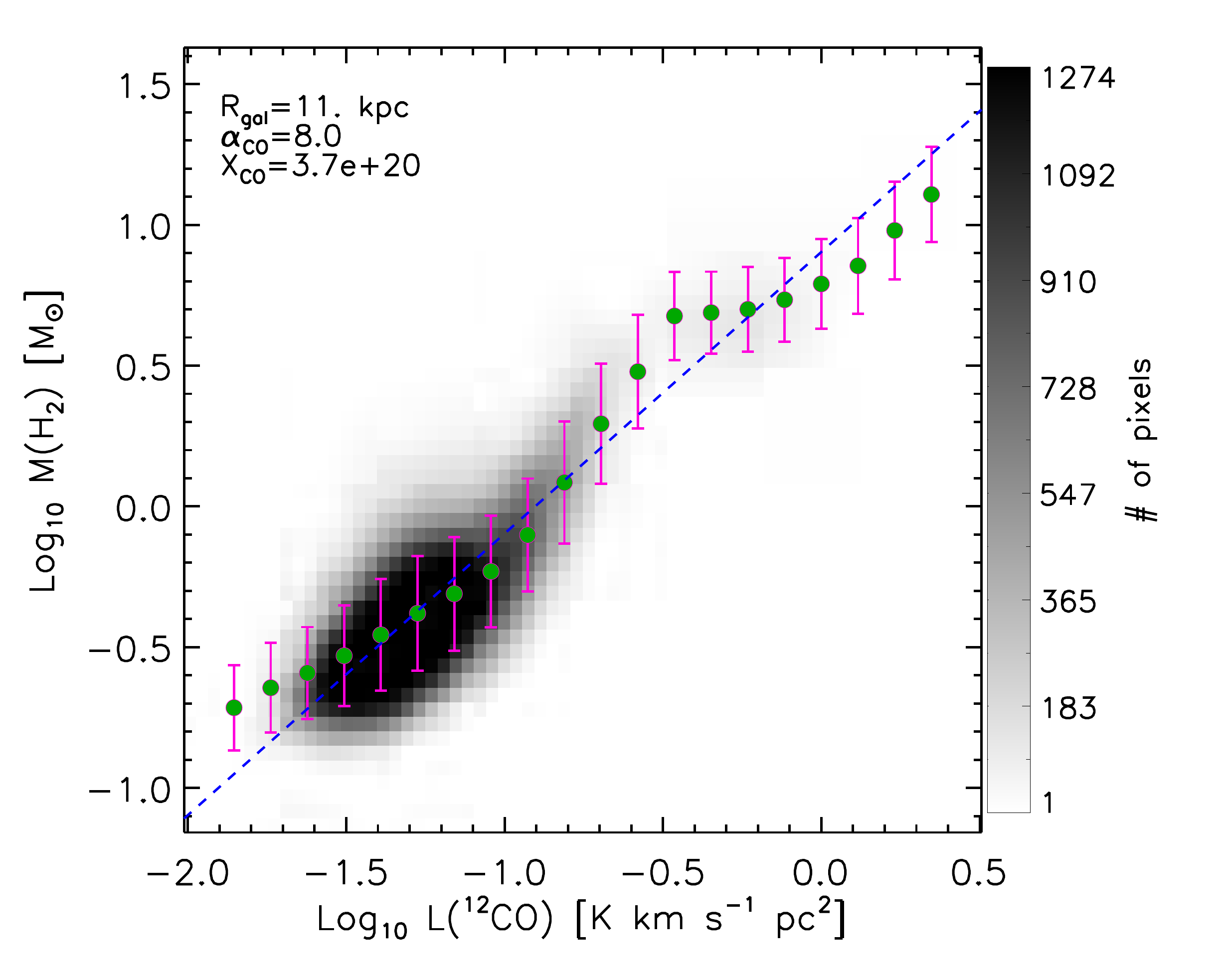} 
             \includegraphics[width=8cm]{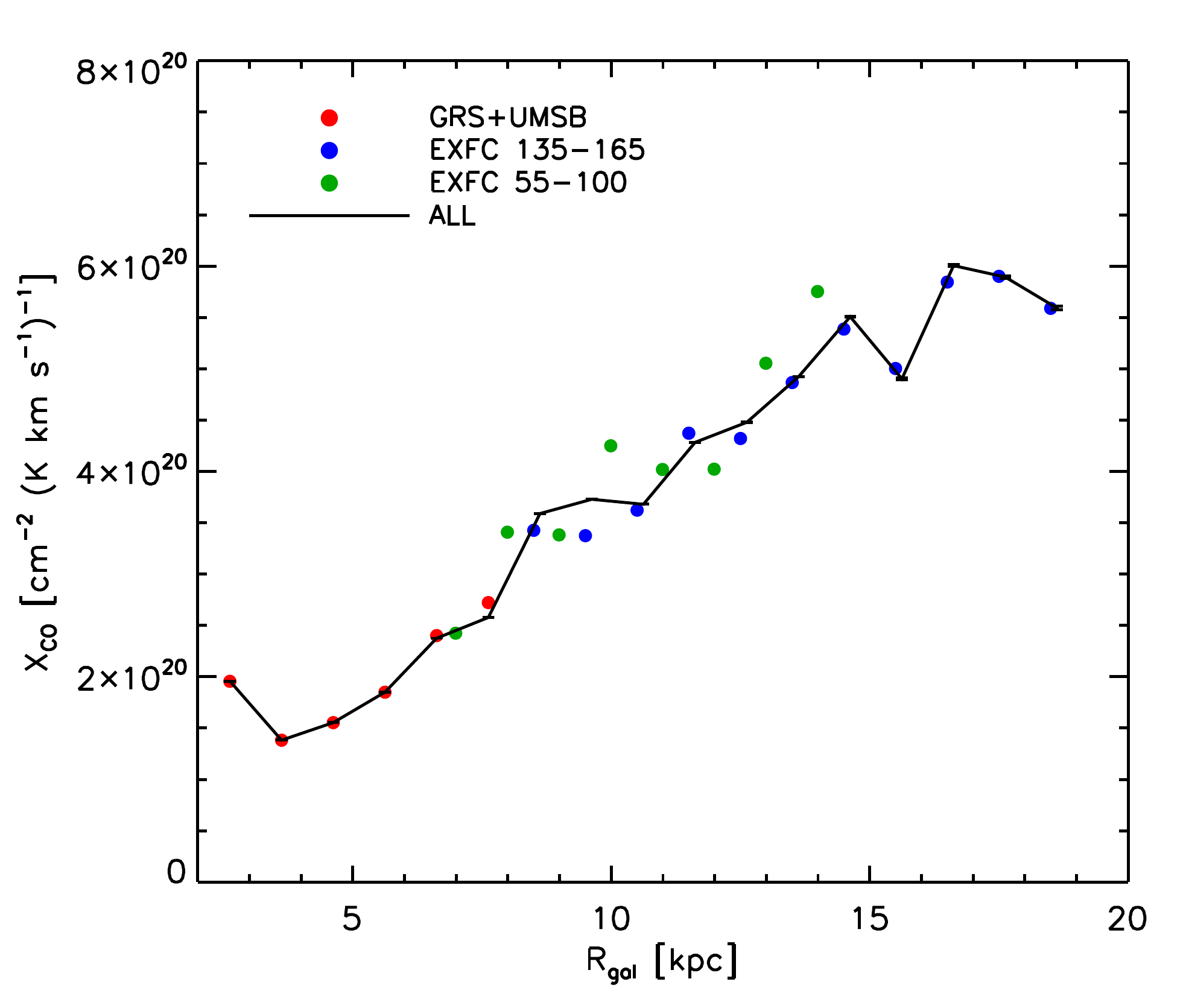} 
	
                              \caption{Relation between the \COTs luminosity of a voxel and its H$_2$ mass at a galactocentric radius $R_{\mathrm{gal}}$ $=$ 5.6 kpc (top) and $R_{\mathrm{gal}}$ $=$ 11 kpc (middle), using the combined data sets (GRS+UMSB, EXFC 55-100 and 135-165). The relations are derived from voxels with \COs emission $>$2$\sigma_{13}$ only. The slope of the relation is indicated in the legend, and corresponds to $X_{CO}$ $=$ 1.9$\times 10^{20}$ cm$^{-2}$ (K km s$^{-1}$)$^{-1}$ s at $R_{\mathrm{gal}}$ $=$ 5.6 kpc , and $X_{CO}$ $=$ 3.7$\times 10^{20}$ cm$^{-2}$ (K km s$^{-1}$)$^{-1}$ at $R_{\mathrm{gal}}$ $=$ 11 kpc. The bottom panel shows the $X_{CO}$ factor as a function of Galactocentric radius $R_{\mathrm{gal}}$. }
\label{lum_mass_rel}
\end{figure}

\indent For each voxel with \COTs detection, the \COTs excitation temperature $T_{\mathrm{ex}}$ is computed following Equation 1 of \citet{romanduval2010}. For each voxel with both \COTs and \COs detections, the  \COs optical depth $\tau_{13}$ is also computed using Equation 2 of \citet{romanduval2010}. Using the distances to each voxel (with a unique solution in the outer Galaxy, and 10 realizations of the near/far ambiguity in the inner Galaxy) and Equation 9 of \citet{romanduval2010}, we derive the \COTs and \COs luminosities in all voxels with detections, as well as the H$_2$ mass M(H$_2$) in voxels with \COTs and \COs detections. \citet{romanduval2010} used a constant abundance ratio of 45 between \COTs and \COs in order to convert the \COs optical depth of a mass of H$_2$. In this work, which includes a much larger range in galactocentric radius, we adopt the abundance ratio derived in \citet{milam05}, which is characterized by a radial gradient:

\begin{equation}
\frac{n(^{12}CO)}{n(^{13}CO)} = 6.2 \times R_{gal} + 18.7
\end{equation}

\indent In voxels with \COs main beam temperatures $>$2$\sigma_{13}$, there is a tight linear relation between the \COTs luminosity of a voxel and its H$_2$ mass, as derived from \COTs and \CO. The slope of this relation is the CO-to-H$_2$ conversion factor, $X_{\mathrm{CO}}$ (for column density) or $\alpha_{\mathrm{CO}}$ (for surface density), and increases with increasing Galactocentric radius, $R_{\mathrm{gal}}$. This relation between $L$(\COT) and $M$(H$_2$), derived from the combined data sets, is shown for $R_{\mathrm{gal}}$ $=$ 5.6 kpc and 11 kpc in the top two panels of Figure \ref{lum_mass_rel}. The bottom panel of Figure \ref{lum_mass_rel} displays the variations of the CO-to-H$_2$  conversion factor with Galactocentric radius.  $X_{\mathrm{CO}}$ varies between 1.5$\times 10^{20}$ cm$^{-2}$ (K km s$^{-1}$)$^{-1}$ at $R_{\mathrm{gal}}\sim 3$ kpc and 6$\times 10^{20}$ cm$^{-2}$ (K km s$^{-1}$)$^{-1}$  at $R_{\mathrm{gal}}\sim 15$ kpc. This is in agreement with the conclusions in \citet{goldsmith08}, who found that the mass of both diffuse and dense CO-emitting gas in the Taurus molecular cloud  is well traced by its luminosity, albeit with a slightly lower conversion factor of 4.1 \Msu pc$^{-2}$ (K km s$^{-1}$)$^{-1}$, corresponding to $X_{\mathrm{CO}}$ $=$ 2$\times 10^{20}$ cm$^{-2}$ (K km s$^{-1}$)$^{-1}$. \citet{liszt10} also reached similar conclusions in a study of diffuse Galactic sight-lines.\\ 
\indent In Figure \ref{lum_mass_rel}, we fit a linear relation, thus forcing the slope in log-log space to be 1. However, to investigate potential deviations from a linear relation, we also performed a linear fit in log-log space (power-law fit), leaving the slope as a free parameter. The resulting slopes were between 0.96 and 1.001, indicating that the relation between CO luminosity and H$_2$ mass is well described by a linear function. 
\indent Since we derive detection masks from smoothed data, the \COs and \COTs main beam temperatures of voxels within the detection mask can be smaller than the uncertainties, or even negative. Instances of this effect are visible in the wings of the CO lines in Figure \ref{example_masking}. While necessary to avoid thresholding-induced positive biases (as described in Section \ref{det_noise_section}), this effect creates some numerical issues in the computation of $\tau_{13}$ and  M(H$_2$). To circumvent this issue, the M(H$_2$) values in voxels with \COs main beam temperatures lower than 2$\sigma_{13}$ are replaced with estimates derived from the relation between the \COTs luminosity of a voxel and its mass. Since the CO-H$_2$ relation depends on Galactocentric radius, we bin the data in radial intervals of width 1 kpc, and derive a CO-to-H$_2$ conversion factor in each radial bin from the voxels with \COs detections $>2\sigma_{13}$. We then apply the same conversion factor between \COTs luminosity and H$_2$ mass to the voxels in that same radial bin, but with \COs main beam temperatures $<2\sigma_{13}$.  This  not only allows us to derive an H$_2$ mass for voxels in the dense mask, albeit with \COs below the 2$\sigma$ sensitivity, but also to derive an H$_2$ mass in the diffuse CO component, where \COs is not detected and a mass estimate would otherwise not be possible. 

\section{Results}\label{results_section}

\subsection{General properties and filling factor of the diffuse, dense, and very dense CO gas}

\indent Tables 2 and 3 summarize the number of voxels in the ``noise'' and ``detection'' categories for each line (\COT, \CO, CS), and the number of ``diffuse'', ``dense'', and ``very dense'' voxels in each survey. In Tables 2 and 3, we also list the corresponding filling factors, computed as the number of detected voxels divided by the total number of voxels in the PPV cubes. CO and CS emission are in general relatively sparse in the PPV cubes, particularly in the outer Galaxy. The \COTs line has the highest filling factor in the PPV cubes, with 24\% of voxels in the ``detection'' mask in the inner Galaxy covered by the GRS+UMSB. The filling factor in the PPV cubes of the \COTs line drops to $<$2\% in the outer Galaxy. The filling factors in the PPV cubes of the \COs line is about half that of the \COTs line.  The CS-emitting gas is much more compact with a filling factor of $<$5\%. This progression is qualitatively seen in Figures \ref{grs_maps} and \ref{cs_maps}. Within the \COT-emitting gas, the diffuse \COTs gas fills a slightly smaller volume (39\%) than the dense CO gas (61\%) in the GRS+UMSB coverage, but diffuse and dense CO gas occupy equal volumes in the outer Galaxy.\\

\begin{figure}
   \centering
               \includegraphics[width=8cm]{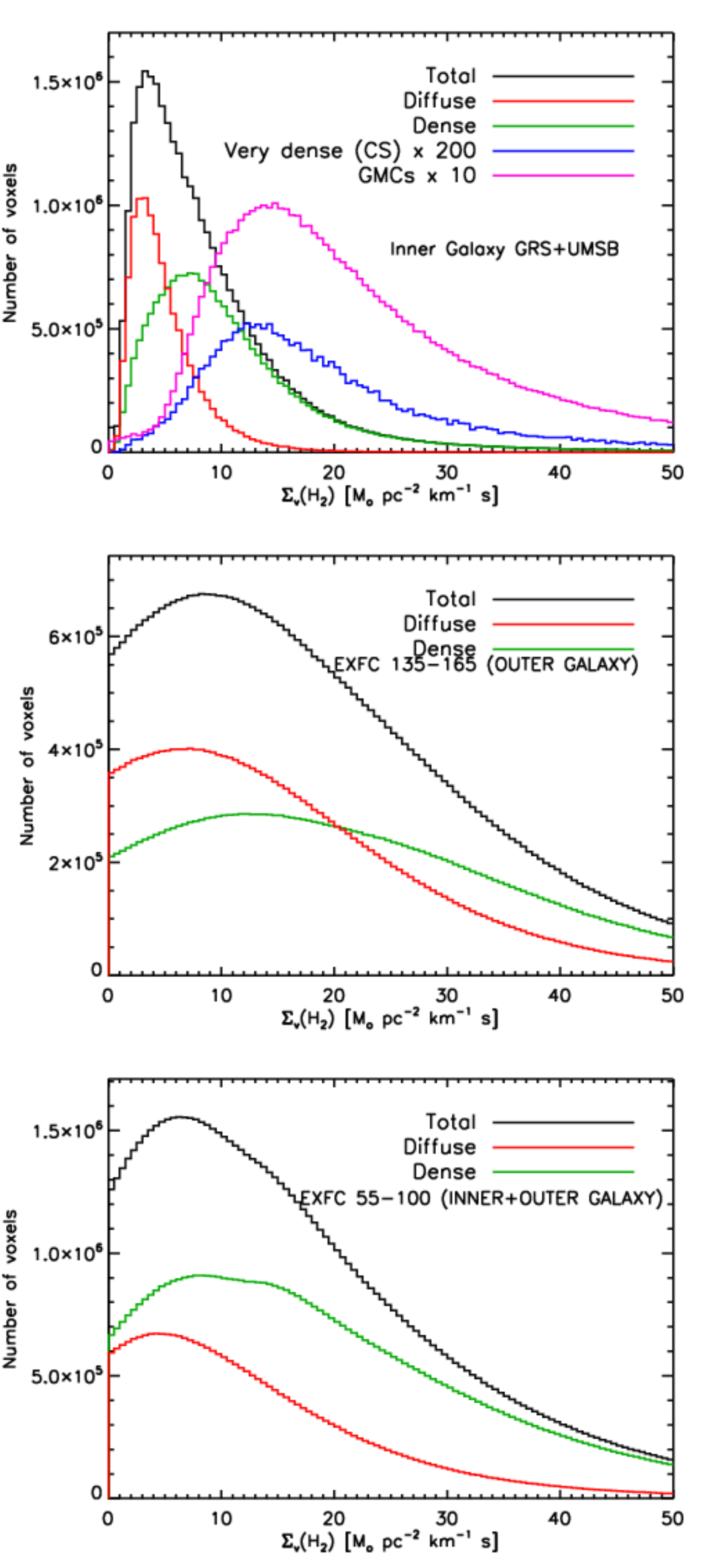} 
       \caption{Distribution of the H$_2$ spectral surface density in each voxel (surface density along the line-of-sight per unit velocity, expressed in \Msu pc$^{-2}$ (km s$^{-1}$)$^{-1}$) in each survey. The top panel corresponds to the inner Galaxy with the GRS+USMB surveys. The middle panel includes data from the EXFC 135-165 survey (outer Galaxy), and the bottom panels shows data from the EXFC 55-100 survey (inner and outer Galaxy). The diffuse (detected in \COT, undetected in \CO), dense (detected in \COTs and \CO), and very dense (detected in \COT, \CO, and CS, inner Galaxy only) components are indicated by red, green, and blue lines, respectively. The black line corresponds to the total contribution of these 3 components. The distribution of spectral surface densities in voxels located within giant molecular clouds identified in the GRS by \citet{romanduval2010} is shown in magenta. In the inner Galaxy, the field where CS observations were obtained only covered 2 deg$^2$ and therefore corresponds to a number of voxels too small to be seen in the histogram. We thus plot the number of voxels in this ``very dense'' category multiplied by $\sim$200 so that it can be clearly seen in the histogram plots. }
\label{plot_histos_mh2}
\end{figure}

\indent Figure \ref{plot_histos_mh2} shows the distribution of the ``spectral surface density of H$_2$'' in each voxel, $\Sigma_v$(H$_2$), expressed in \Msu pc$^{-2}$ (km s$^{-1}$)$^{-1}$, for the diffuse, dense, and very dense components, in each survey. Spectral surface densities correspond to the surface density of H$_2$ along the line-of-sight at the velocity of the voxel, per unit velocity. As expected, the distributions of the diffuse, dense, and very dense components peak at increasingly higher spectral surface densities (3, 8 and 12 \Msu pc$^{-2}$ (km s$^{-1}$)$^{-1}$ respectively). We note that, in order to obtain the surface density of a parcel of molecular gas, the spectral surface densities need to be multiplied by the velocity width of that parcel. Hence, the spectral surface densities shown in Figure \ref{plot_histos_mh2} cannot be directly compared to the typical surface densities of GMCs. For comparison however, we also show in Figure \ref{plot_histos_mh2} the distribution of the spectral surface densities in voxels within the GMCs identified in \citet{romanduval2010}, which closely resembles the spectral surface density distribution of the very dense gas, peaking at  $\Sigma_v$(H$_2$) $=$15 \Msu pc$^{-2}$ and exhibiting a long tail to high spectral surface densities.   \\
\indent The distribution of the H$_2$ spectral surface density in each voxel in the EXFC survey appears wider. However, the difference in the width of the distribution between the GRS+UMSB and EXFC is most likely due to the difference in measurement errors ($\sigma_{12}$) at the original resolution of the data (typically $\sigma_{12}$ $\sim$ 2K per native (22.5''$ \times$ 22.5''$\times$0.13 \kmsn) voxel in EXFC, compared to $\sigma_{12}$ $\sim$ 0.24---0.5 K per native (1'$ \times$1'$\times$0.3 \kmsn) voxel in the GRS+UMSB). \\

\subsection{Spatial distribution of molecular gas in the Milky Way: a view from above}

\begin{figure*}
   \centering
            \includegraphics[width=5.5cm]{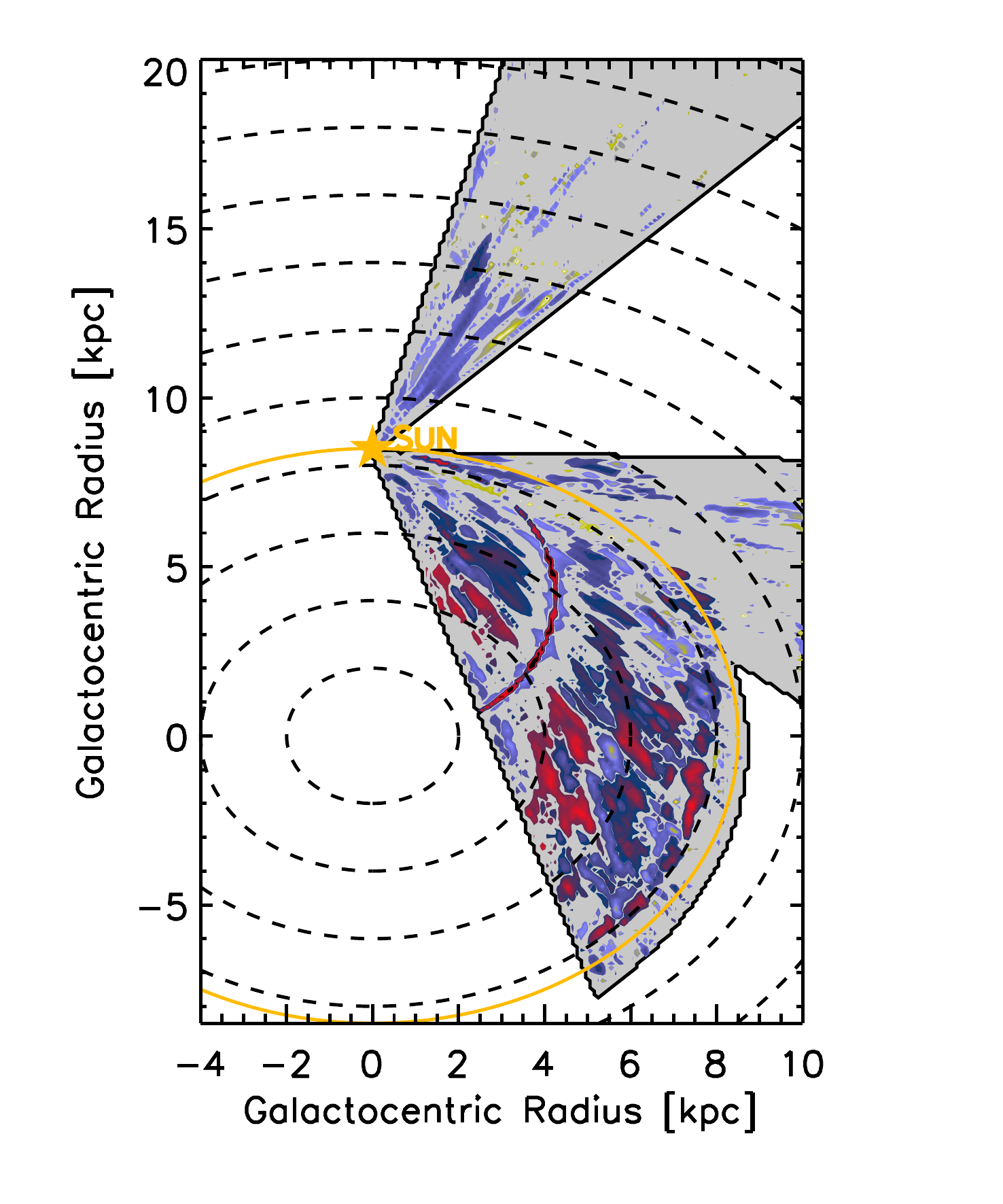} 
                        \includegraphics[width=5.5cm]{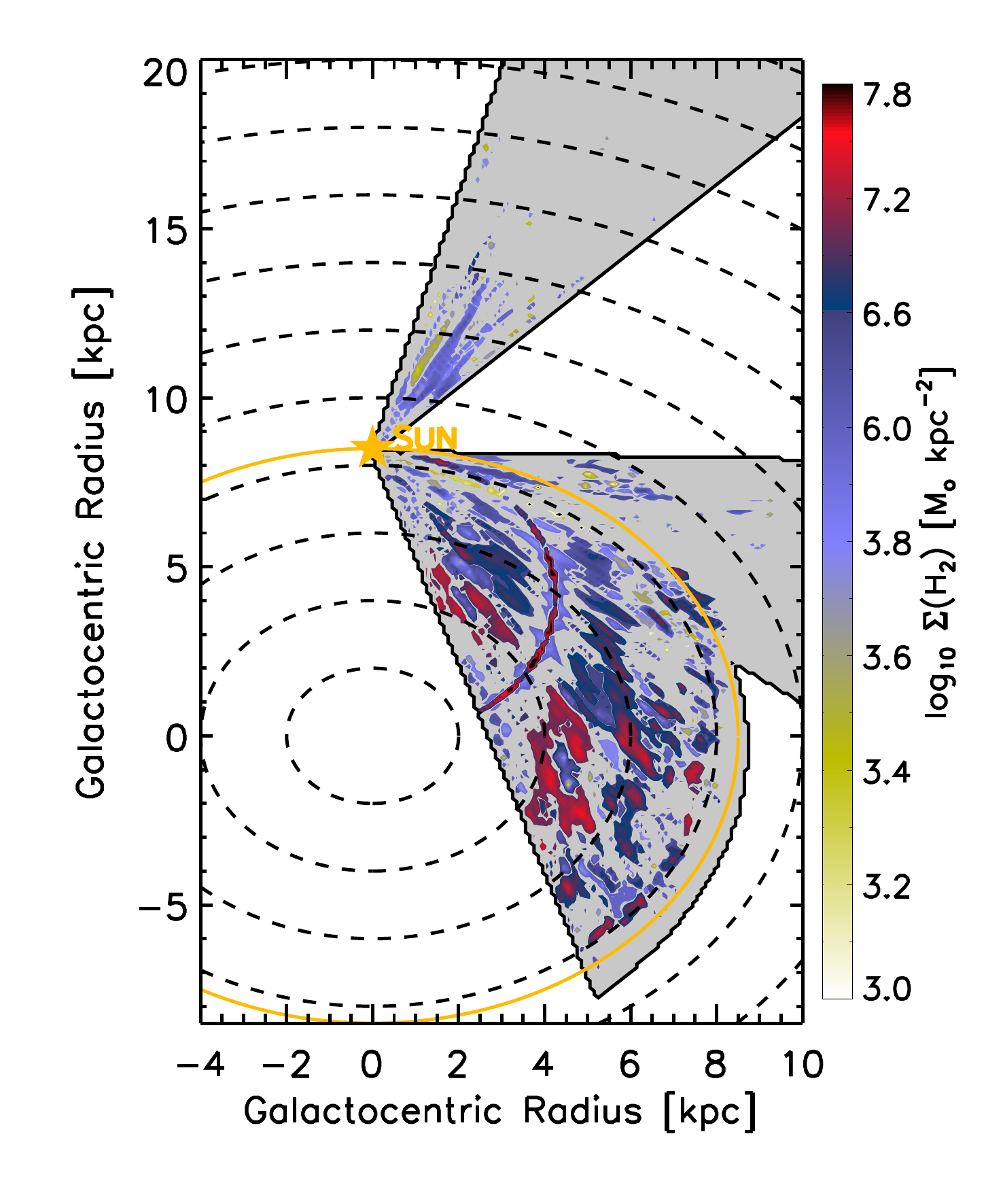} 
                         \includegraphics[width=5.5cm]{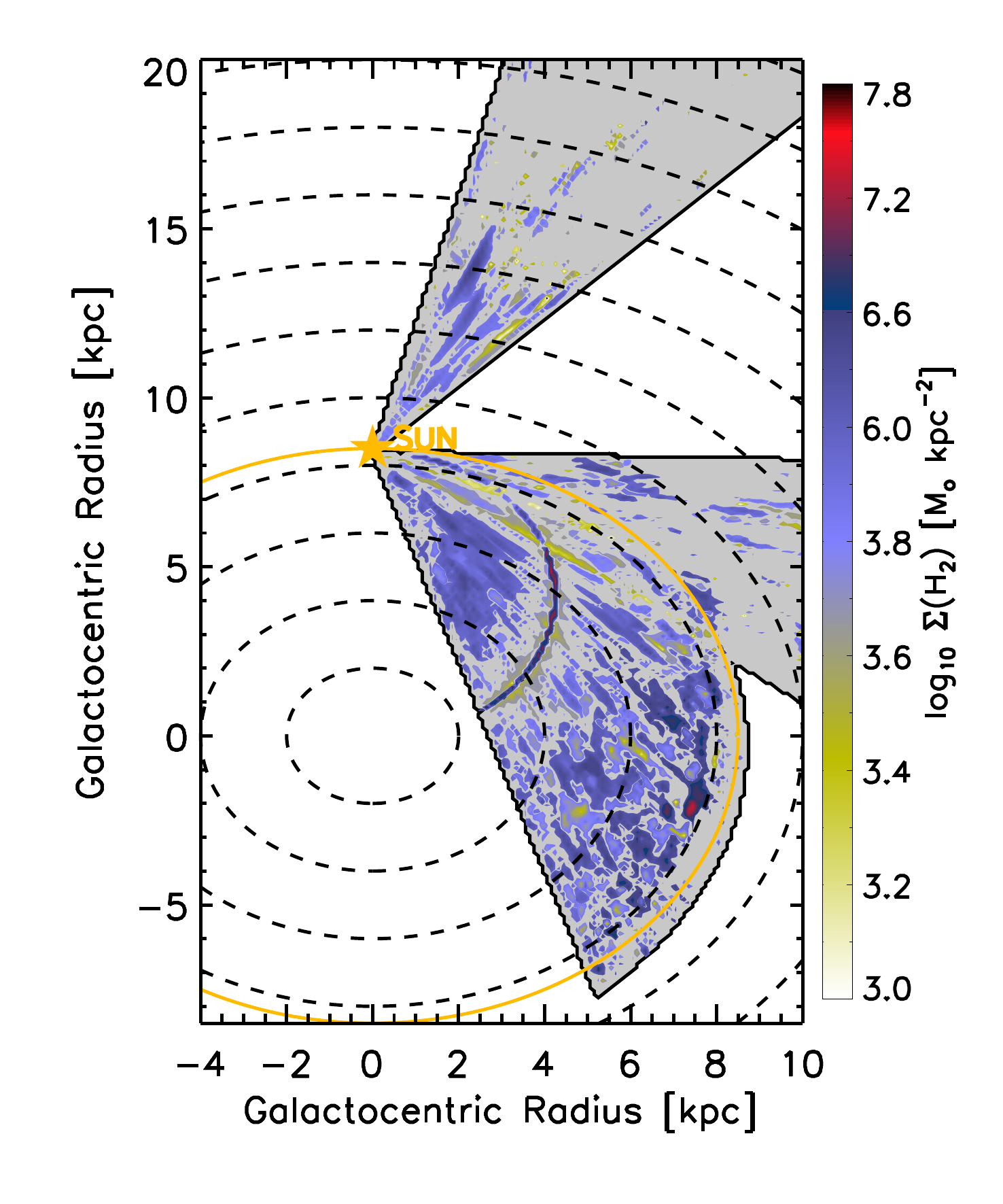} 

       \caption{Map of the total (left), dense (middle, detected in \COTs and \CO), and diffuse (right, detected in \COT, undetected in \CO) average Galactic surface density of molecular gas in the Milky Way. The coverage of the two surveys is indicated by the grey/white contrast. }
       \label{maps_from_above}
\end{figure*}

\indent With the knowledge of the location (distance and coordinates), luminosity and mass of each voxel, we have produced a face-on map of the Galactic distribution of \COT-emitting molecular gas, separating the diffuse (\COT-bright, \CO-dark) and dense (\COT-bright, \CO-bright) CO components. The maps were obtained by summing in each 100 pc $\times$ 100 pc pixel the masses of all voxels located within a pixel, and dividing by the area of that pixel. The resulting face-on maps are shown in Figure \ref{maps_from_above}. Strikingly, the surface density of molecular gas decreases by 1---2 orders of magnitude between Galactocentric radii of 3 kpc and 15 kpc. In both sides of the solar circle, the diffuse CO component is smoother and more uniform than the dense component, which is consistent with the conclusions of \citet{pety13} in M51, who found that 50\% of the CO luminosity in M51 originates from kpc-scale diffuse emission. \\
\indent The large uncertainties on the distance will undoubtedly affect the detailed spatial distribution of molecular gas in the face-on maps. These maps should therefore not be used to derive the detailed structure of the Milky Way, but rather are meant to better conceptualize and visualize the transformations involved, between looking {\it through} the Galactic Plane and {\it from above} the Galactic Plane. In the next section, the radial and vertical distributions of H$_2$ and of the different CO components are computed by averaging those maps in bins of Galactocentric radius and vertical height above the plane.

\subsection{Radial distribution of the \COT, \COs and CS average Galactic integrated intensities}

\begin{figure*}
   \centering
            \includegraphics[width=8cm]{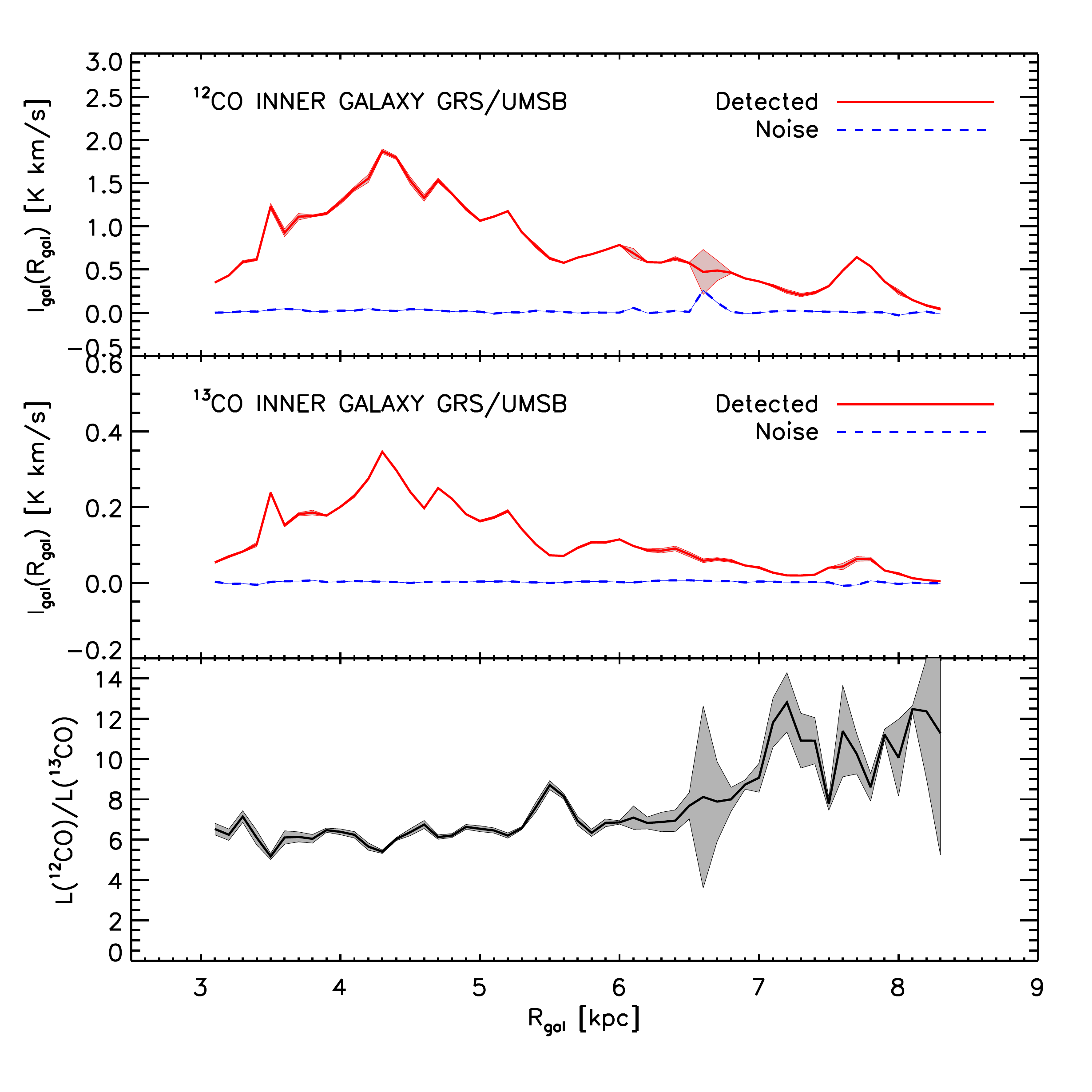} 
            \includegraphics[width=8cm]{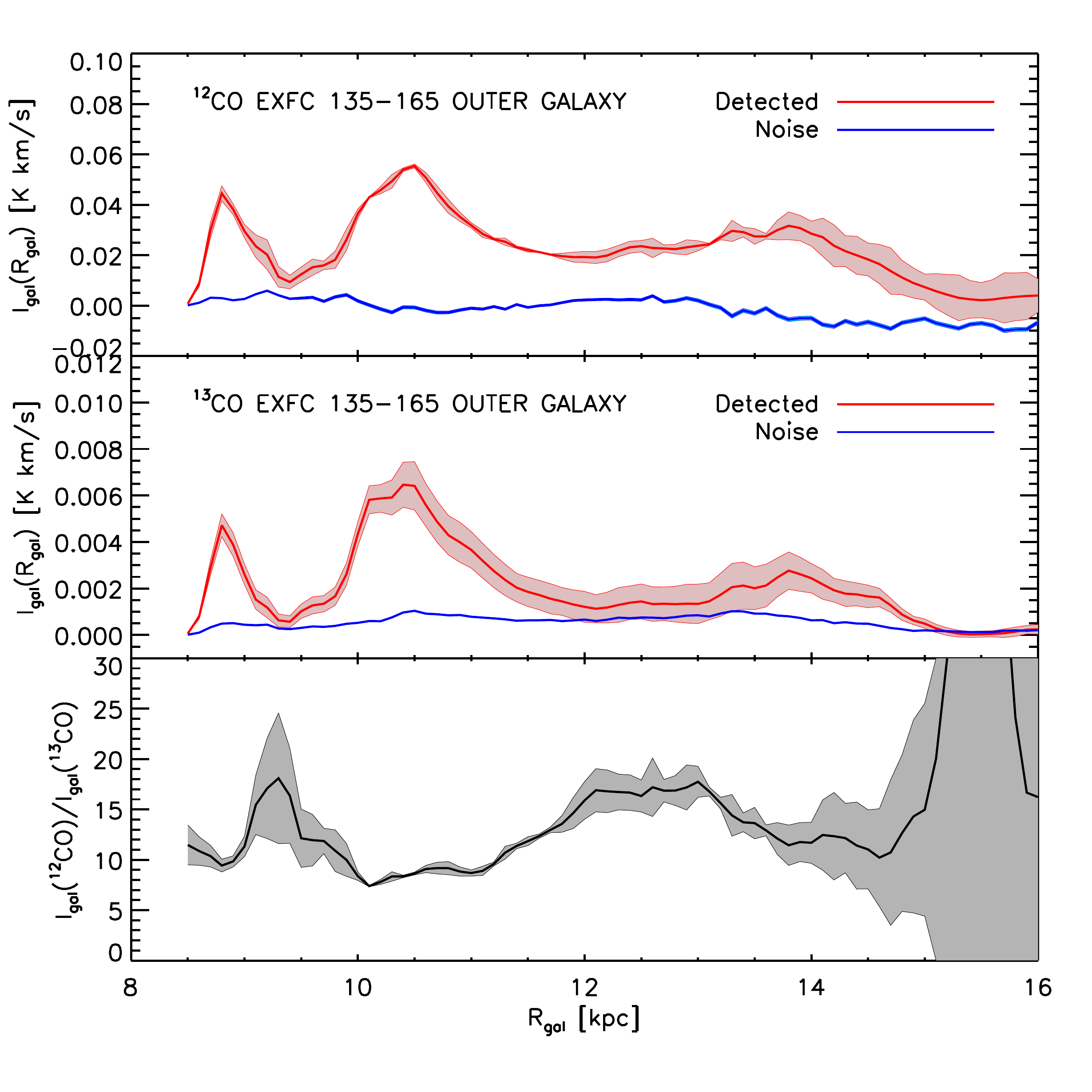} 
            \includegraphics[width=8cm]{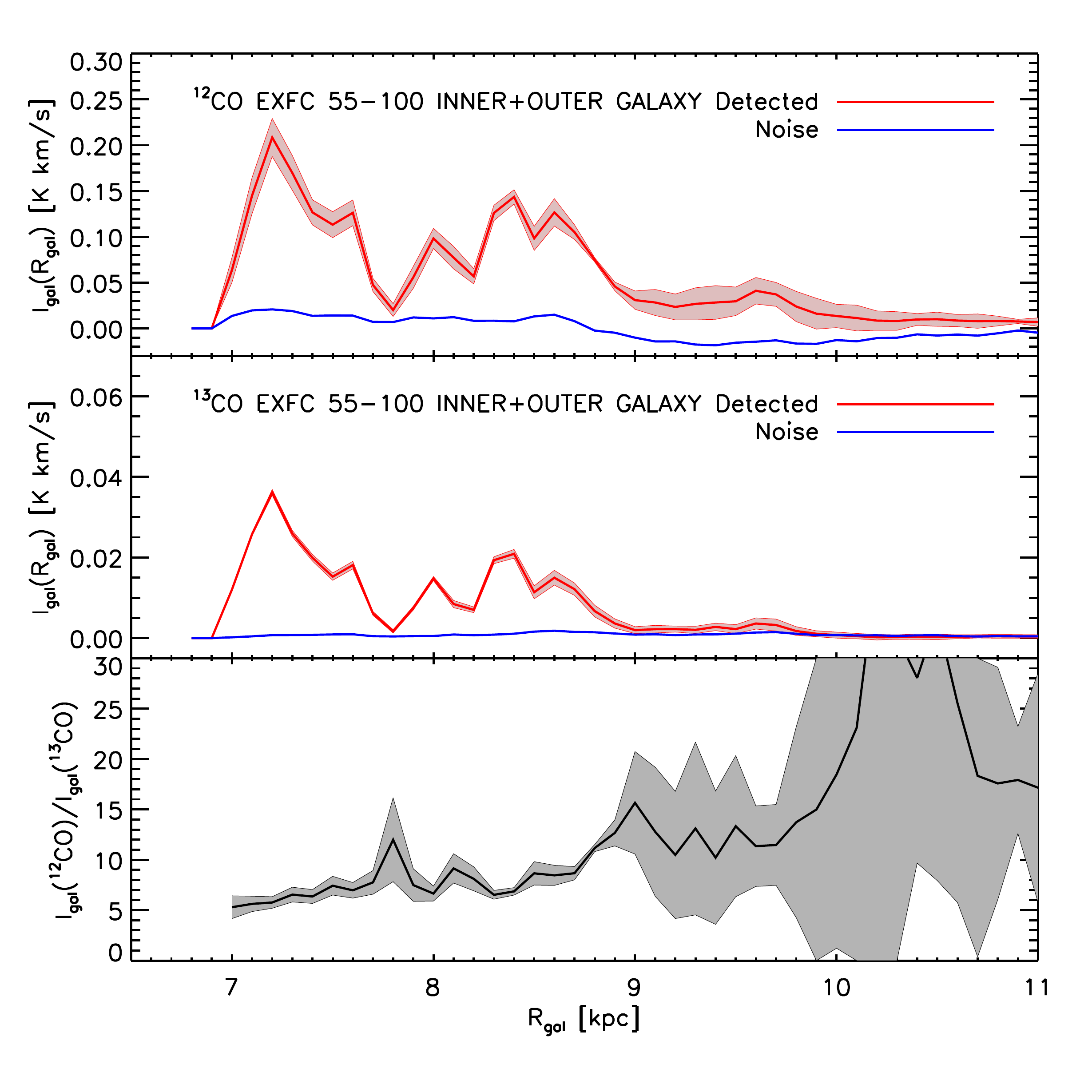} 
                          \caption{Detected (red) and noise (blue) \COTs and \COs average Galactic integrated intensities as a function of Galactocentric radius (in bins of width 0.1 kpc), in the GRS+UMSB surveys(top left), in EXFC 135-165 (top right) and in EXFC 55-100 (bottom). The ratio of the \COTs and \CO intensities is shown in the bottom sub-panels.}
\label{plot_rgal_tot_lco}
\end{figure*} 

\begin{figure}
   \centering
            \includegraphics[width=8cm]{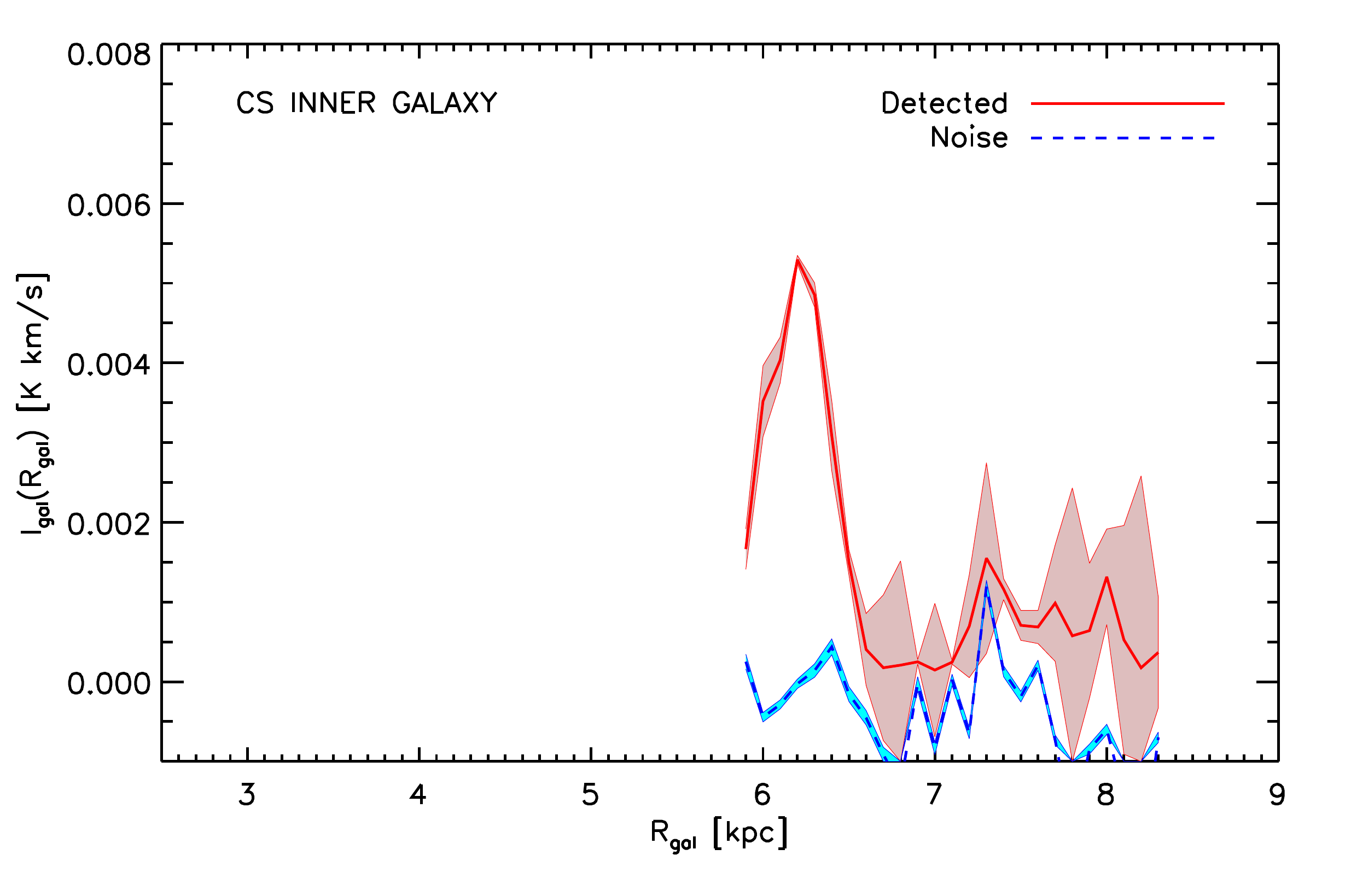} 
                                              
                          \caption{Detected (red) and noise (blue) CS average Galactic integrated intensities as a function of Galactocentric radius (in bins of width 0.1 kpc), in the 2 deg$^2$ field of the GRS}
\label{plot_rgal_tot_lcs}
\end{figure} 

\indent We first examine the distribution of \COT, \CO, and CS average Galactic integrated intensities $I_{\mathrm{gal}}$ with Galactocentric radius $R_{\mathrm{gal}}$. Here, average Galactic integrated intensity corresponds to the total luminosity in a Galactocentric radius bin, divided by the surface area covered by the survey in that bin, projected onto the Galactic disk. $I_{\mathrm{gal}}$ is therefore the integrated intensity as seen from above the plane, averaged over radial bins of width 0.1 kpc, and is equivalent to the integrated intensity measurements for extragalactic surveys of face-on galaxies. In the inner Galaxy, the average Galactic integrated intensities are obtained for each Monte-Carlo realization of distances separately. We then average the trends of $I_{\mathrm{gal}}(R_{\mathrm{gal}})$ versus $R_{\mathrm{gal}}$ over all ten realizations, and we include the standard deviation between realizations in the final error estimation. As an additional check that our detection algorithm picks up all the low level extended emission, we also compute the average Galactic integrated intensities of the voxels with non-detections (``noise voxels') using a similar procedure. The resulting average Galactic integrated intensities for \COTs and \CO, as well as their ratio, are plotted as a function of galactocentric radius in Figure \ref{plot_rgal_tot_lco} for each survey. The average Galactic integrated intensity in the CS 2-1 line is plotted versus $R_{gal}$ in Figure \ref{plot_rgal_tot_lcs}. Of course, these trends are representative of the gas seen within the coverage of the surveys. \\
\indent To compute the total error on $I_{\mathrm{gal}}(R_{\mathrm{gal}})$, indicated at 1$\sigma$ by the thickness of the curves in Figures \ref{plot_rgal_tot_lco} and  \ref{plot_rgal_tot_lcs}, we sum in quadrature the different sources of errors. These sources include errors on the near and far distance estimation due to non-circular motions, for the inner Galaxy, the standard deviation between Monte-Carlo distance realizations, and the residuals from the average Galactic integrated intensity of ``noise'' voxels. For a given voxel, the error on its luminosity incurred by the error on its distance is given by $\delta L = 2L \delta d/d$, where $d$ and $\delta d$ are the distance of the voxel and its error and $L$ its luminosity. The error on the total luminosity in a Galactocentric radius bin is the quadratic sum of the errors on the luminosities of all the voxels included in that bin. We note that the standard deviation between near/far distance realizations in the inner Galaxy is negligible compared to the other sources of errors. \\
\indent The average Galactic integrated intensity of \COTs and \COs decreases by one to two orders of magnitude between $R_{\mathrm{gal}}$ $\sim$ 3 kpc and  $R_{\mathrm{gal}}$ $\sim$ 15 kpc.  $I_{\mathrm{gal}}(R_{\mathrm{gal}})$ for \COTs and \COs track each other closely throughout the Galactic plane, with an approximately constant ratio of 5 out to $R_{\mathrm{gal}}$ $=$ 6.5 kpc. The \COT/\COs integrated intensity ratio increases to 10---12 in the solar neighborhood, although the errors are larger in this case, and remains between 10 and 20 in the outer Galaxy, out to $R_{\mathrm{gal}}$ $=$ 14 kpc. The \COT/\COs integrated intensity ratio appears to be anti-correlated with $I_{\mathrm{gal}}(R_{\mathrm{gal}})$, or in other words, with average Galactic surface density of CO-bright molecular gas. This factor of 2 increase in the \COT/\COs luminosity ratio between 3 kpc and the solar neighborhood has previously been observed by \citet{liszt84}. They interpret it as being a result of the volume density decreasing away from the Galactic center, which would be consistent with the decreasing star formation rate and resulting cloud temperatures seen in \citet{romanduval2010}. Additionally, the radial trend in the \COT/\COs luminosity ratio could be explained by the decrease in the fraction of dense gas with decreasing surface density (and increasing Galactocentric radius), and the \CO/\COTs abundance gradient observed in \citet{milam05}, which varies between 50 at $R_{gal}$ $=$ 5 kpc and 100 at  $R_{gal}$ $=$ 15 kpc and could be consistent with the variations in luminosity ratio.  Other possible effects that could explain the variations in the \COT/\COs integrated intensity or luminosity ratio include more sub-thermally excited \COs in the outer Galaxy.  \\

\subsection{Radial distribution of the diffuse, dense, and very dense CO components}\label{radial_section}

 \begin{figure*}
   \centering 
                          \includegraphics[width=12cm]{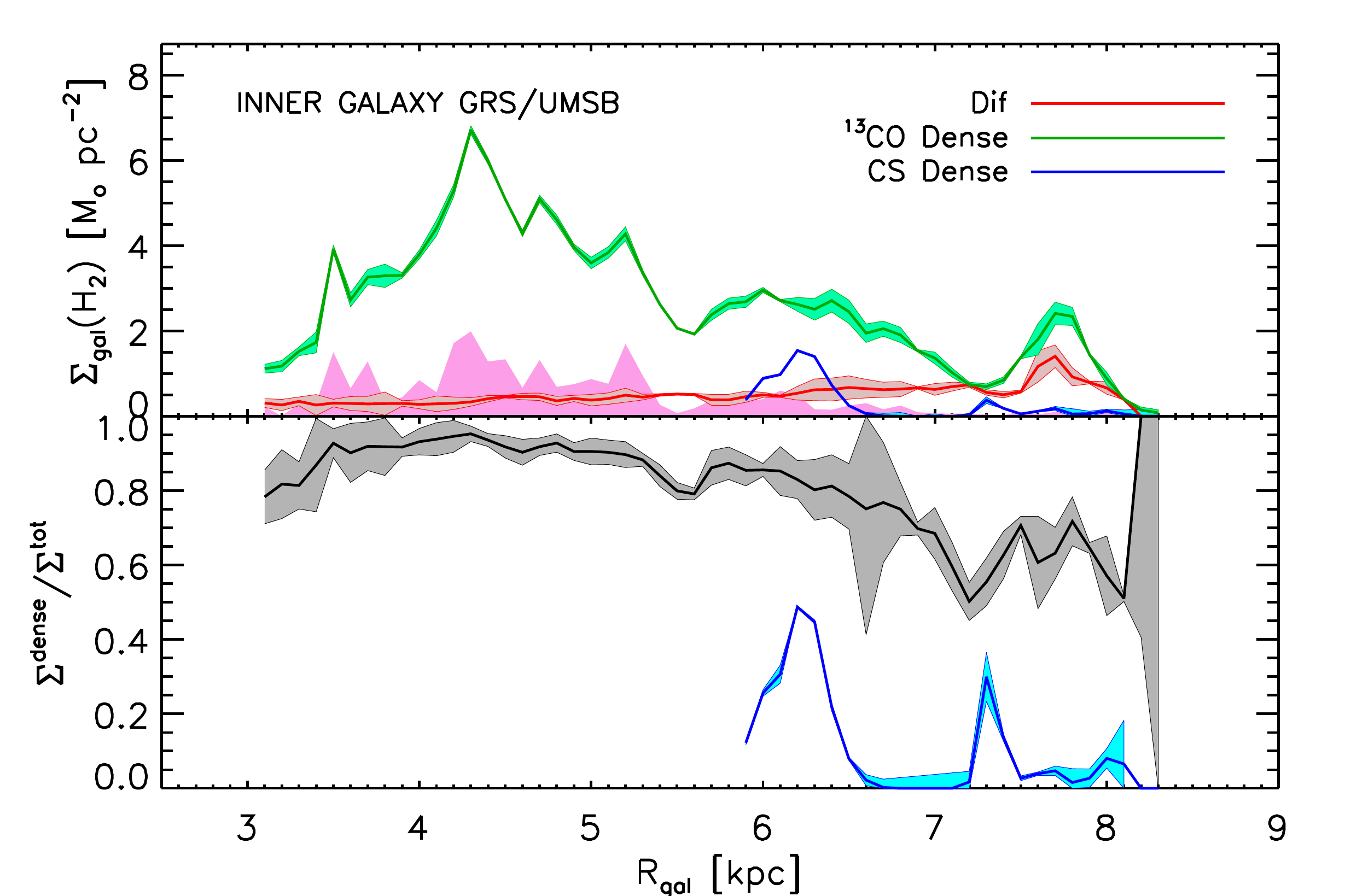} 
                           \includegraphics[width=12cm]{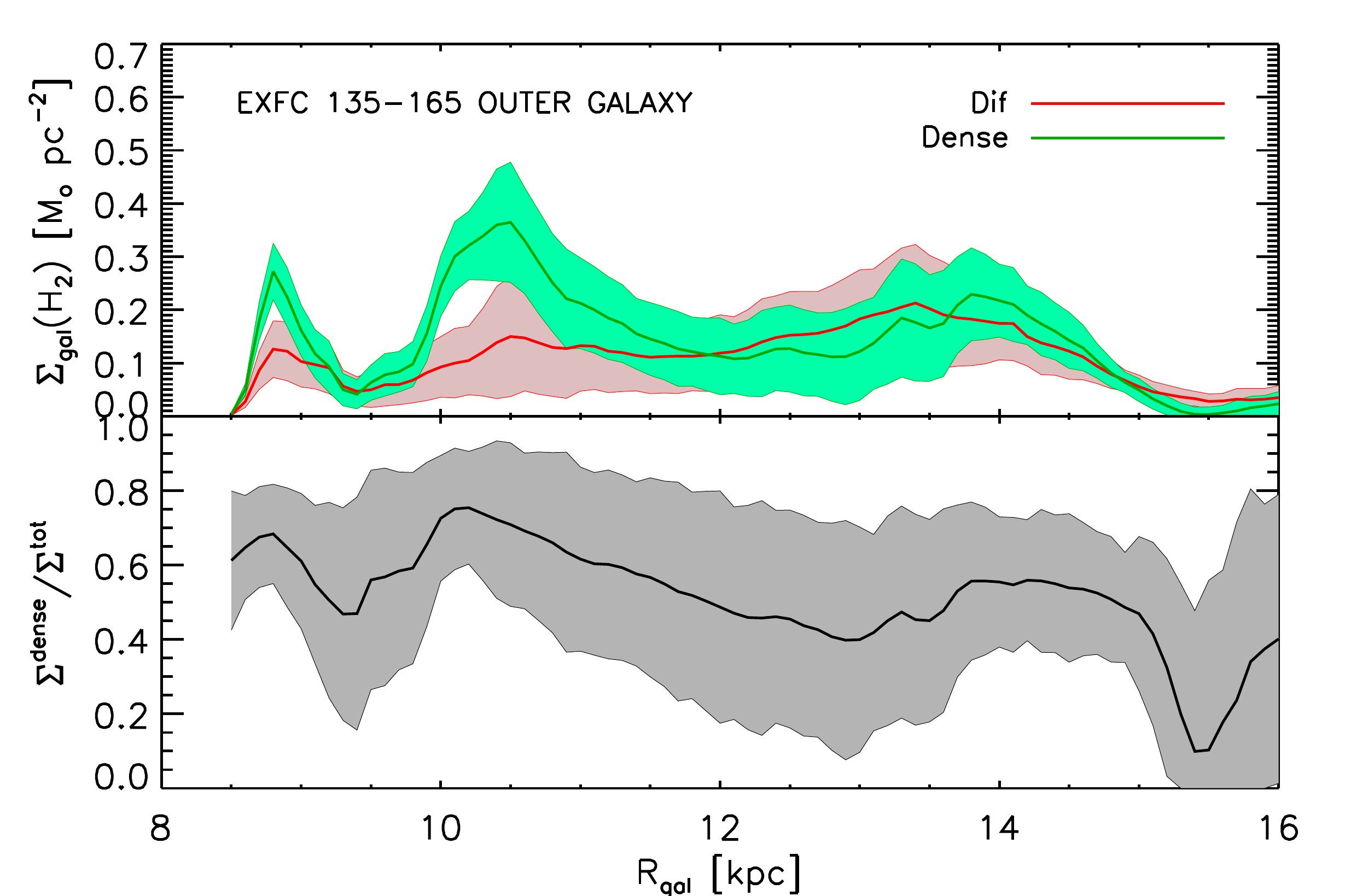} 
                           \includegraphics[width=12cm]{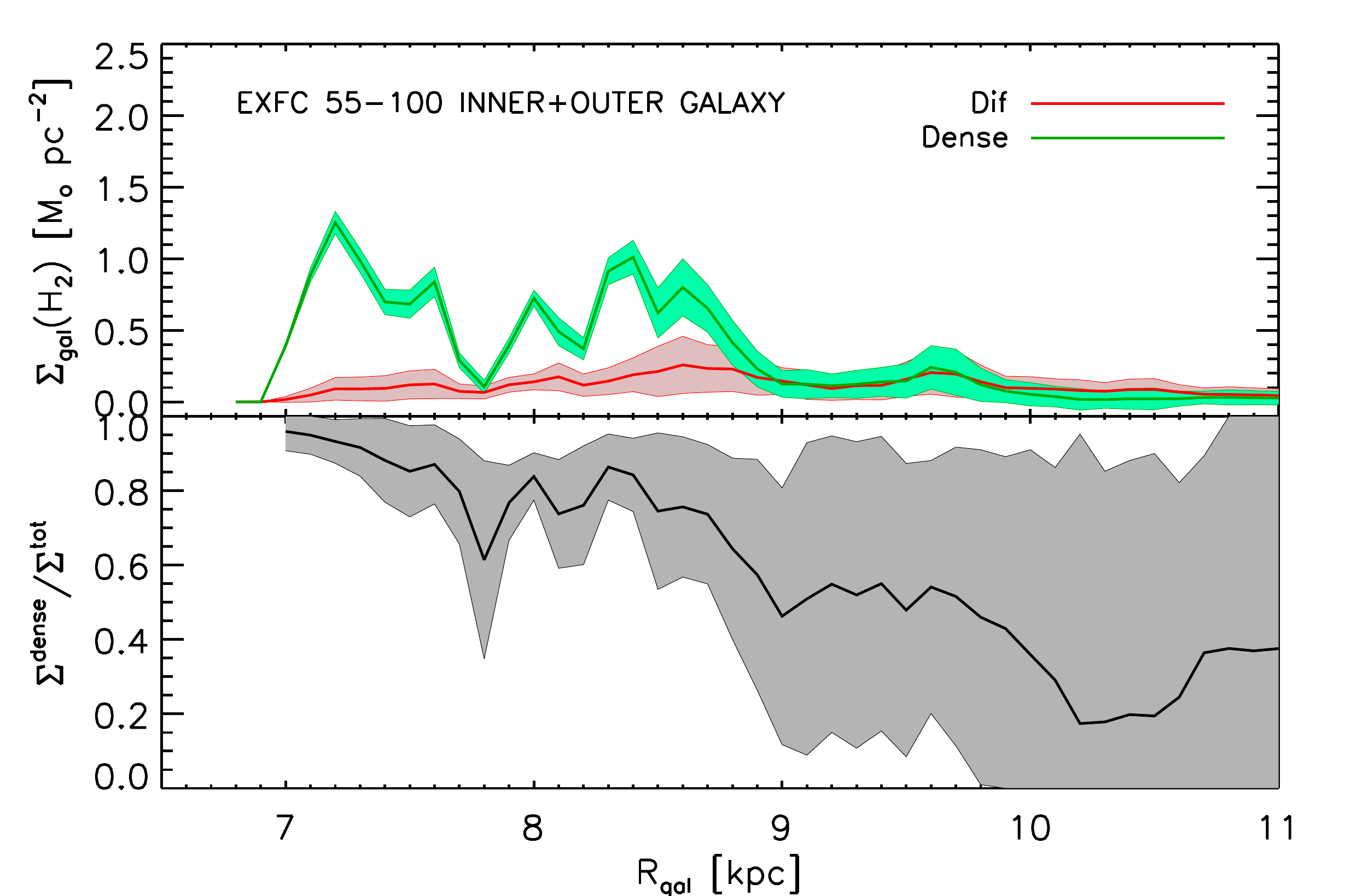}

                            \caption{Average Galactic H$_2$ surface densities of the diffuse (red, detected in \COT, undetected in \CO), dense (green, detected in \COTs and \CO), and very dense (blue, detected in \COT, \CO, and CS) components  averaged in bins of width 0.1 kpc, as a function of Galactocentric radius in the GRS+UMSB (inner Galaxy only, top), in the EXFC 135-165 survey (outer Galaxy only, middle), and in the EXFC 55-100 survey (inner and outer Galaxy, bottom).  In the inner Milky Way covered by the GRS, the pink filled curve indicates the surface density of H$_2$ in molecular clouds identified with a clump finding algorithm in \citet{romanduval2010}}
\label{plot_rgal_dif_dense_lco}
\end{figure*}

 \begin{figure*}
   \centering 
                          \includegraphics[width=12cm]{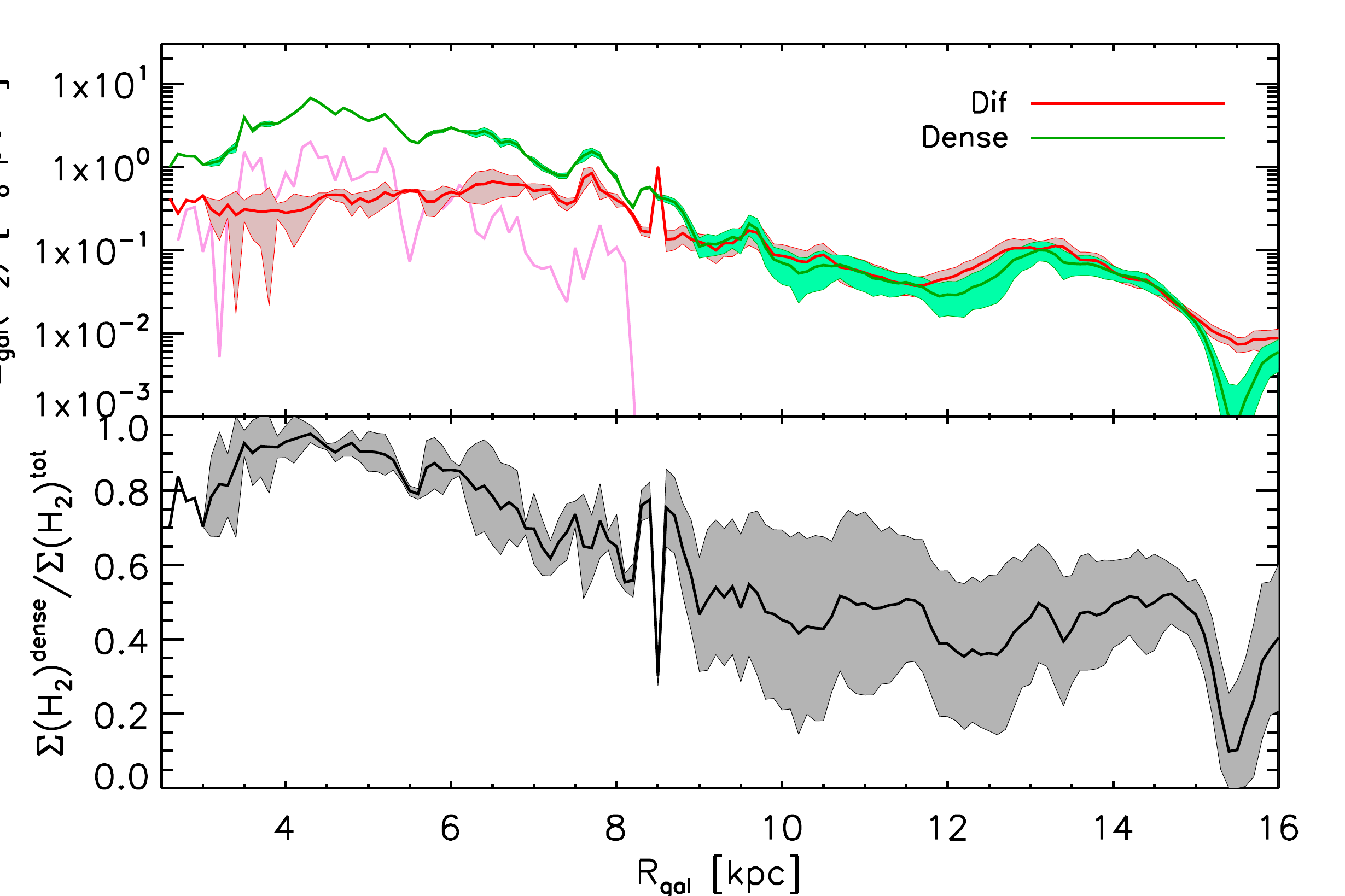} 
                            \caption{Average Galactic H$_2$ surface densities of the diffuse (red, detected in \COT, undetected in \CO) and dense (green, detected in \COTs and \CO) components as a function of Galactocentric radius (in bins of width 0.1 kpc), in logarithmic scale, combining all data sets. In the inner Galaxy, the pink line indicates the surface density of H$_2$ in molecular clouds identified in \citet{romanduval2010}. }
\label{plot_rgal_dif_dense_lco_log}
\end{figure*}

\begin{deluxetable}{ccccc}
\tabletypesize{\scriptsize}
\tablecolumns{5}
\tablewidth{8cm}
\tablecaption{Total luminosity and molecular mass in the Milky Way in the diffuse and dense components traced by \COT.}
\tablenum{5}
 
 \tablehead{
& & Inner & Outer & Total
 }
 
 \hline
 \startdata

\multirow{3}{*}{L($^{12}$CO)} & Diffuse  & 2.0$\times 10^1$ & 4.0 & 2.4$\times 10^1$\\
& Dense & 1.1$\times 10^2$ & 3.8 & 1.1$\times 10^2$\\
& Very dense & 4.8 &  --- & 4.8 \\
 & Total & 1.3$\times 10^2$ & 7.7 & 1.4$\times 10^2$\\
 
 &&&&\\
 \hline 
 &&&&\\
 
\multirow{3}{*}{M(H$_2$)} & Diffuse & 9.3$\times 10^7$ & 6.0$\times 10^7$ & 1.5$\times 10^8$\\
& Dense & 4.6$\times 10^8$ & 3.9$\times 10^7$ & 4.9$\times 10^8$\\
& Very dense & 2.9$\times 10^7$ & --- & 2.9$\times 10^7$\\
 & Total & 5.5$\times 10^8$ & 9.9$\times 10^7$ & 6.5$\times 10^8$\\

  \enddata
  \label{total_masses}
  
  \tablecomments{Luminosities are given in units of K km s$^{-1}$ kpc$^2$. Masses are given in \Msun. Statistical errors on integrated luminosities and masses are $\sim$1\%.  Systematic uncertainties are $\sim$30\% due to uncertainties on abundances (\CO/\COTs and \COT/H$_2$) and the possibly non-applicable assumption of LTE in the diffuse regime, which systematically underestimates masses at low column densities.}

%     \hline
\end{deluxetable}

\indent Similarly, we derive the average Galactic H$_2$ surface density $\Sigma_{gal}(\mathrm{H}_2)(R_{\mathrm{gal}})$ in the diffuse extended, dense, and very dense components, by summing the masses of all voxels in each CO-component in Galactocentric radius bins of width 0.1 kpc, and dividing by the surface area of each survey projected on the Galactic Plane. The resulting radial distributions of the three CO gas components are shown in linear space separately for each survey in Figure \ref{plot_rgal_dif_dense_lco} and in logarithmic space combining all data sets in Figure \ref{plot_rgal_dif_dense_lco_log}. As for the \COTs and \COs average Galactic integrated intensity computation, all relevant sources of errors are included in the final error budget. \\
\indent In the inner Galaxy, the dense component dominates in mass. The mass fraction of dense gas decreases from 90\% at $R_{\mathrm{gal}}$ $=$ 4 kpc to 50\% in the solar neighborhood. In the outer Galaxy, the mass fraction of dense gas varies between 40\% and 80\%, and is anti-correlated with surface density. \\
\indent Assuming that the Galaxy is roughly axisymmetric and that the radial trends observed in our surveys are representative of the Galaxy as a whole, the total masses of the diffuse and dense CO components integrated between 3 kpc and 15 kpc are $1.5\times10^8$ \Msu and $4.9\times10^8$ \Msu respectively, or 25\% and 75\% of the total H$_2$ mass traced by \COTs (6.5$\times 10^8$ \Msun) respectively. Statistical errors on integrated masses are $\sim$1\%. Systematic uncertainties are $\sim$30\% due to errors on the assumed abundances (\CO/\COTs and \COT/H$_2$) and the possibly inaccurate assumption of pure LTE at low column densities, which systematically underestimates masses in the diffuse regime. The total mass of H$_2$ derived here is compatible within errors to the number quoted in the review by \citet{heyer15} or 9$\times10^8$ \Msun. The total luminosities and masses of each component, in the inner, outer, and entire Galaxy are listed in Table 5.  \\
 \indent In the 2 deg$^2$ field where CS observations are available at $\ell$ $=$ 45\degn, the very dense CO component (also bright in CS) is sparse. In the Galactocentric radius range probed by the observations ($R_{gal}$ $=$ 6---8.5 kpc), the fraction of the very dense component in the total surface density varies between zero and 50\% locally in presumably massive star formation regions. Locally, the very dense component can thus comprise a significant fraction of the gas. However, the very dense gas has a relatively low filling factor. The very dense component traced by CS has a total mass of 2.9$\times 10^8$ \Msu in the observed range ($R_{gal}$ $=$ 6---8.5 kpc). The total molecular gas mass in this radial interval is 1.8$\times 10^8$ \Msun. Therefore, the very dense component represents only $\sim$14\% of the total molecular gas mass traced by \COTs emission. However, it is possible that the fraction of very dense gas traced by CS be higher closer to the center of the Galaxy. \\
 \indent As a comparison, \citet{battisti14} found that the very dense component of molecular clouds, as traced by mm dust continuum emission, comprises $\sim$10\% of the mass of molecular clouds identified with the CPROPS detection algorithm \citep{rosolowsky06}. In Section \ref{comparison_to_gmcs_section}, we show that about 15\% of the total molecular gas mass traced by \COTs in the Milky Way resides in such GMCs, and so the very dense gas fraction determined in \citet{battisti14} would represent about 1.5\% of the total H$_2$ mass traced by \COT, which is slightly lower than the very dense gas fraction derived here.

\subsection{Anti-correlation between diffuse CO-gas and Galactic surface density of molecular gas}

\begin{figure}
   \centering
            \includegraphics[width=8cm]{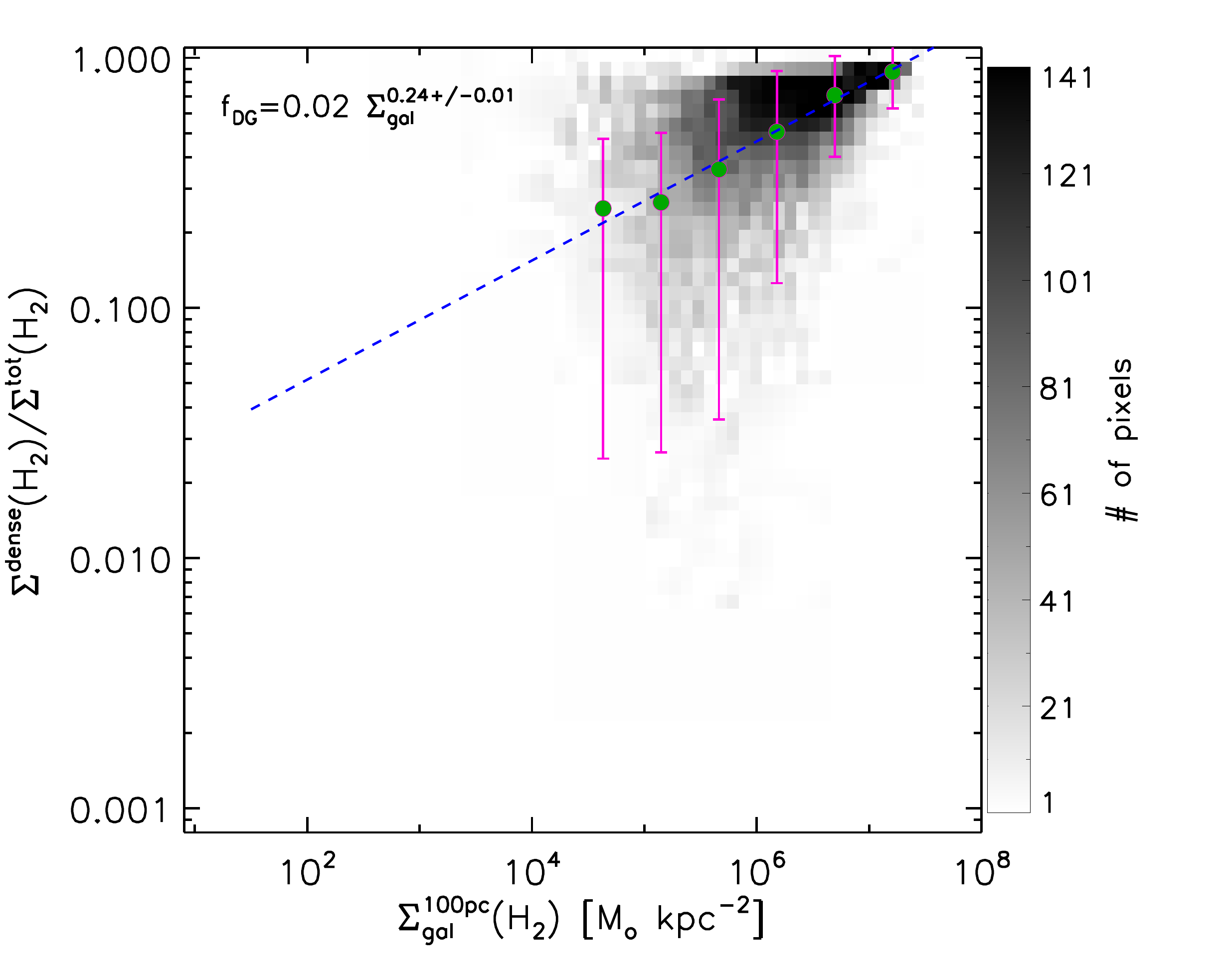}                   
  	\includegraphics[width=8cm]{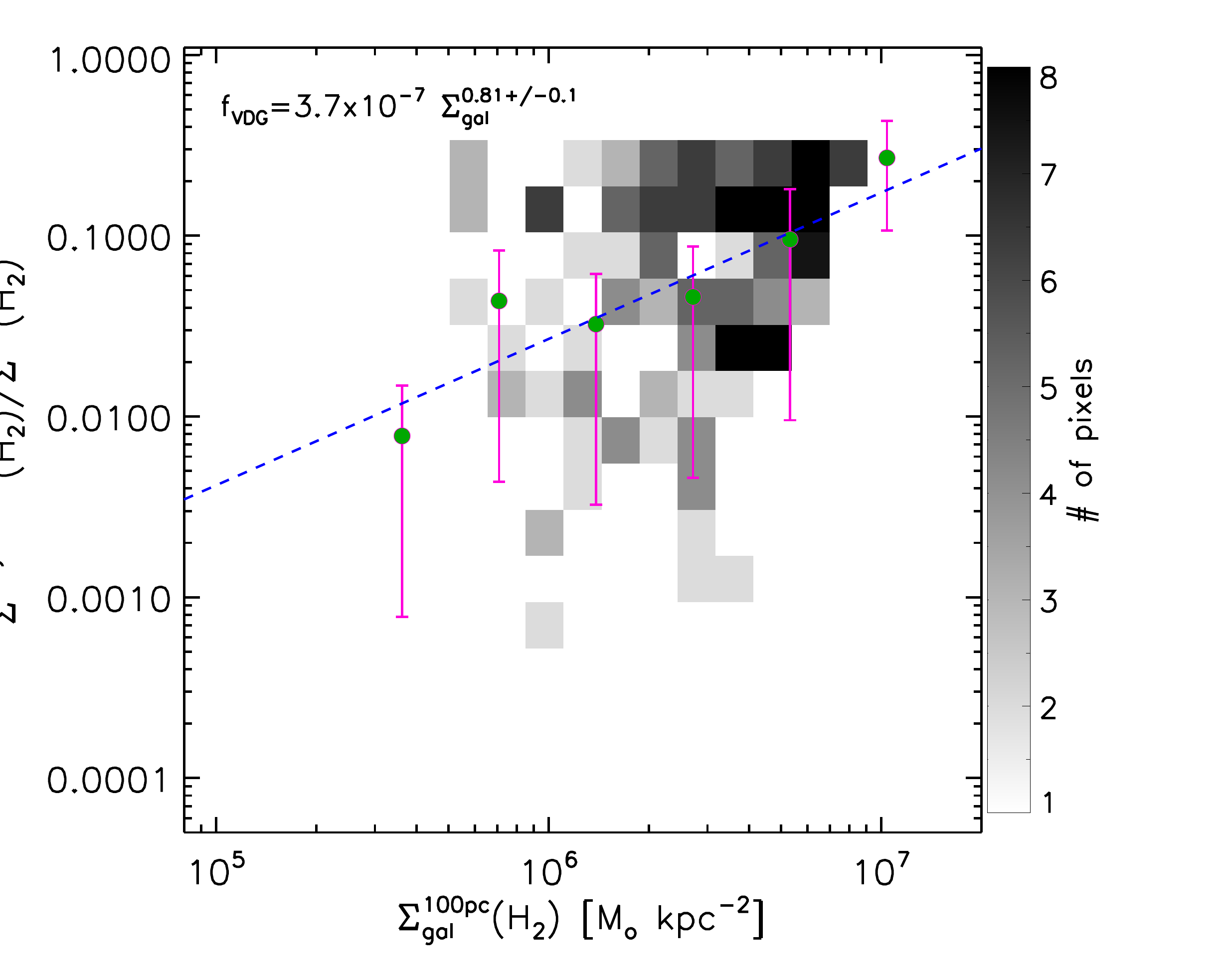}
       \caption{Relation between dense (top, detected in \COTs and \CO) and very dense (bottom, detected in \COT, \CO, and CS) molecular gas fraction and average Galactic surface density of molecular gas, derived from the combined data sets. The grey scale indicates the density of points, while the pink/green dots show the binned average. The errors bars correspond to the standard deviation in each bin. }
\label{dif_frac_sigma}
\end{figure}

\indent  Figure \ref{plot_rgal_dif_dense_lco} suggests that the fraction of dense CO-gas is correlated with the Galactic surface density of molecular gas (traced by \COT). We plot in the top panel of Figure \ref{dif_frac_sigma} the relation between the mass fraction of dense CO gas and the Galactic molecular gas surface density, averaged in 100 pc wide pixels as seen from above the Galaxy (see Figure \ref{maps_from_above}). At low molecular surface densities ($\Sigma_{gal}^{100pc}<10^5$ \Msu kpc$^{-2}$), the fraction of dense gas is low ($f_{DG}<$20---30\%). The dense gas fraction increases when the disk's molecular gas surface density increases, and reaches 80-90\% at high surface densities of $10^7$ \Msu kpc$^{-2}$. A linear fit in log-log space to the $\Sigma_{gal}^{100pc}$---$f_{DG}$ relation yields $f_{DG}$ $=$ 0.02 $\Sigma_{gal}^{0.24\pm0.01}$.\\
\indent The relation between Galactic surface density of H$_2$ and the fraction of very dense gas (traced by CS emission), $f_{VDG}$, is shown in the bottom panel of Figure \ref{dif_frac_sigma}. The mass fraction of very dense gas (presumably star forming) also increases with increasing disk surface density, from $f_{VDG}$ $=$ 1\% at $\Sigma_{gal}^{100pc}$ $=$ a few $10^5$ \Msu kpc$^{-2}$, up to $f_{VDG}$ $=$ 30\% at $\Sigma_{gal}^{100pc}$ $=$ 10$^7$ \Msu kpc$^{-2}$. A linear fit in log-log space yields $f_{VDG}$ $=$ 3.7$\times 10^{-7}$ $\Sigma_{gal}^{0.8\pm0.1}$.\\
\indent

\subsection{Vertical distribution of CO gas in the Milky Way}\label{vertical_section}

\begin{figure*}
   \centering
            \includegraphics[width=\textwidth]{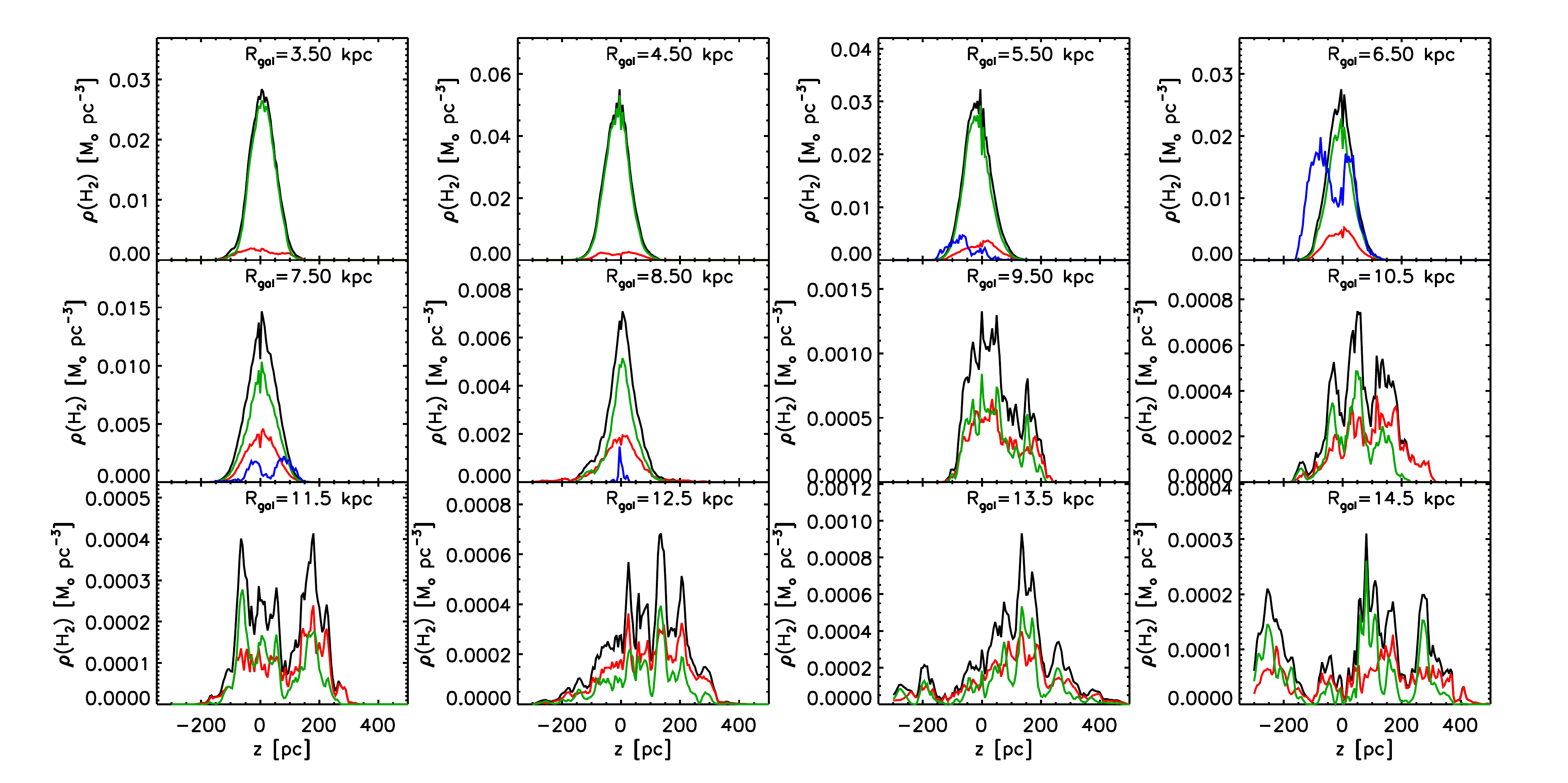} 

                          \caption{Vertical distribution of molecular gas traced by CO as a function of Galactocentric radius, with the contributions of the diffuse extended (detected in \COTs but not \CO) and dense (detected in \COTs and \CO) components in red and green respectively. The vertical distributions are derived with the combined data sets.  in the inner Galaxy, the very dense component traced by CS emission is shown in blue. The black lines corresponds to the total profiles.}
\label{plot_z_distrib_all}
\end{figure*} 

\begin{figure}
   \centering
            \includegraphics[width=8cm]{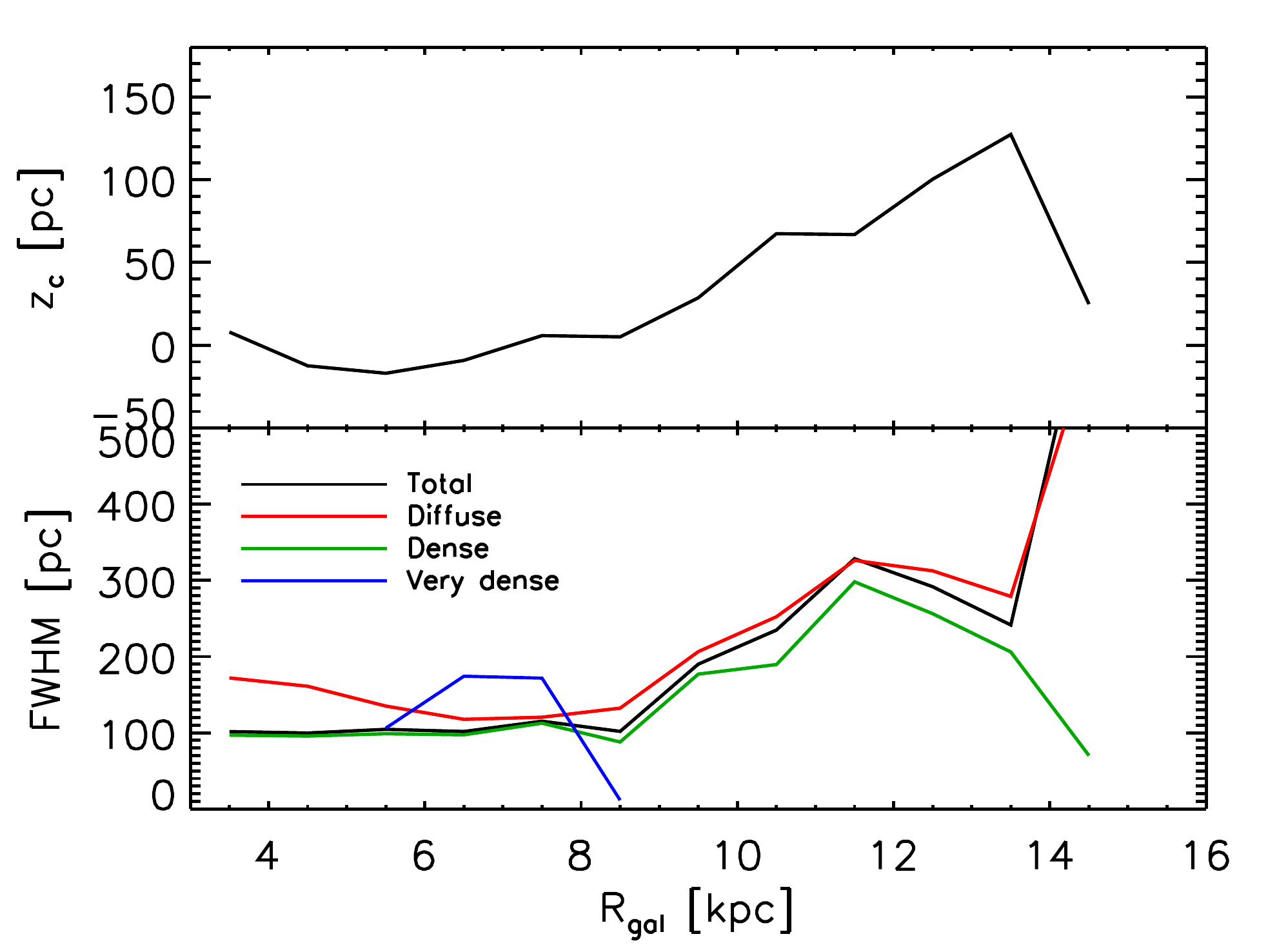} 

                          \caption{Centroid (top) and FWHM (bottom) of the vertical profiles of molecular gas traced by CO as a function of Galactocentric radius, obtained from fitting the vertical profiles shown in Figure \ref{plot_z_distrib_all} to Gaussians. The contributions of the diffuse extended and dense components are shown in red and green respectively. in the inner Galaxy, the very dense component traced by CS emission is shown in blue. The black curve corresponds to the total profile.}
\label{plot_z_distrib_fits}
\end{figure}

\indent We derive the vertical distribution (i.e., perpendicular to the Galactic plane) of the total, diffuse, dense and very dense CO components. Knowing the distance of each voxel, its height above the plane $z$ was computed as $z=d \tan(b)$. We then summed the masses of all voxels in vertical height bins of width 5 pc and galactocentric radius bins of width 1 kpc, divided by the surface areas on the Galactic Plane covered by each survey in those radial bins, and divided by the bin width (5 pc) to obtain the average molecular gas density $\rho(\mathrm{H}_2)$ (in \Msu pc$^{-3}$) as a function of $z$ and $R_{gal}$ for the overall CO gas, as well as the diffuse, dense, and very dense CO gas components. In the inner Galaxy, the vertical profile of the molecular gas density was derived for each Monte-Carlo realization and then averaged over all realizations. The resulting vertical profiles are shown in Figure \ref{plot_z_distrib_all}. The profiles are fitted with Gaussians, and the resulting centroid and FWHM values are plotted as a function of Galactocentric radius in Figure \ref{plot_z_distrib_fits}.  \\
\indent The total vertical profile of molecular gas in the inner Galaxy is well described by a Gaussian function, with a FWHM of $\sim$110 pc. As seen in the radial distribution of diffuse and dense CO gas, the dense CO component dominates the inner Galaxy in mass. The profile of the diffuse component in the inner Galaxy is also Gaussian, but, with a larger FWHM of 130-200 pc.  In contrast, the very dense component is concentrated in the Galactic plane, with a non-Gaussian, double peaked profile of FWHM $\sim$50 pc. \\
\indent In the outer Galaxy, the molecular disk is more warped, with a centroid increasing from a few pc at the solar circle, up to 150 pc at $R_{gal}$ $=$ 14 kpc. The molecular disk is wider than in the inner Galaxy, with FWHM varying between 110 pc and 300 pc. The vertical profiles have multiple peaks and are thus not well fit by a Gaussian. The FWHM shown in Figure \ref{plot_z_distrib_fits} thus represents a gross approximation of the profile width. In the outer Galaxy, the diffuse CO component has a similar mass as the dense CO gas, but their vertical profiles differ significantly. The vertical profile of the diffuse CO gas appears smoother and wider than the profile of the dense CO gas. \\
\indent In both the inner and outer Galaxy, these results are consistent with previous estimates of the thickness and mid-plane displacement summarized by \citet{heyer15}. The larger vertical extent of the diffuse CO component suggests that is originates from a thick disk, which has already been suggested in the Milky Way by \citet{dame94}, and in M51 by \citet{pety13}.

\section{Discussion, limitations, and implications}\label{discussion_section}

\subsection{Comparison to the radial and vertical distribution of molecular gas identified as part of GMCs}\label{comparison_to_gmcs_section}
\indent Studies of the properties and distribution of molecular gas in Galaxies commonly resort to cloud identification algorithm, such as CLUMPFIND \citep{williams94} or dendrograms \citep{rosolowsky08} algorithms. These procedures allow catalogs of discrete objects and associated properties to be derived, including a distance derivation, which cannot be unambiguously determined in the inner Galaxy on a per voxel basis. However, it is not clear what fraction of the total CO emission this type of algorithm picks up. In the left panel of Figure \ref{plot_rgal_dif_dense_lco},  we show the radial distribution of H$_2$ within molecular clouds identified in \citet{romanduval2010}, within the same survey coverage. The molecular gas traced by clouds identified with CLUMPFIND in \citet{romanduval2010} represents a small fraction of the total molecular gas in the inner Milky Way. The total mass of molecular gas in GMCs in the UMSB+GRS coverage is $4.6\times10^7$\Msun, while in this analysis we derive a total molecular gas mass of $3.4\times10^8$\Msu within the same coverage (not to be confused with the mass extrapolated to the entire galaxy in Section \ref{radial_section}, or 6.5$\times10^8$ \Msun). Thus, only $\sim$14\% of the molecular gas mass in the Milky Way was identified within GMCs in the inner Milky Way based on their \COs emission. This number is significantly smaller than the 40\% quoted in \citet{solomon89} and \citet{williams97}. However, these studies identified the GMCs in the \COTs cubes, whereas \citet{romanduval2010} identified CO clouds in the GRS \COs cubes. It is well known \citep[e.g., ][ and this work]{goldsmith08, heyer09} that \COTs emission is more (approximately a factor 2) spatially extended than its \COs counterpart, and so it is not surprising that the mass fraction of CO gas in \COT-identified GMCs is larger than the mass fraction of CO gas in \CO-identified GMCs.  While the Milky Way is more confused than external galaxies, this suggest that studies of molecular gas relying on molecular cloud identification algorithms may be missing the majority of the molecular gas mass.

\subsection{Nature of the ``diffuse'', ``dense'', and ``very dense'' gas}
\begin{figure}
   \centering
            \includegraphics[width=8cm]{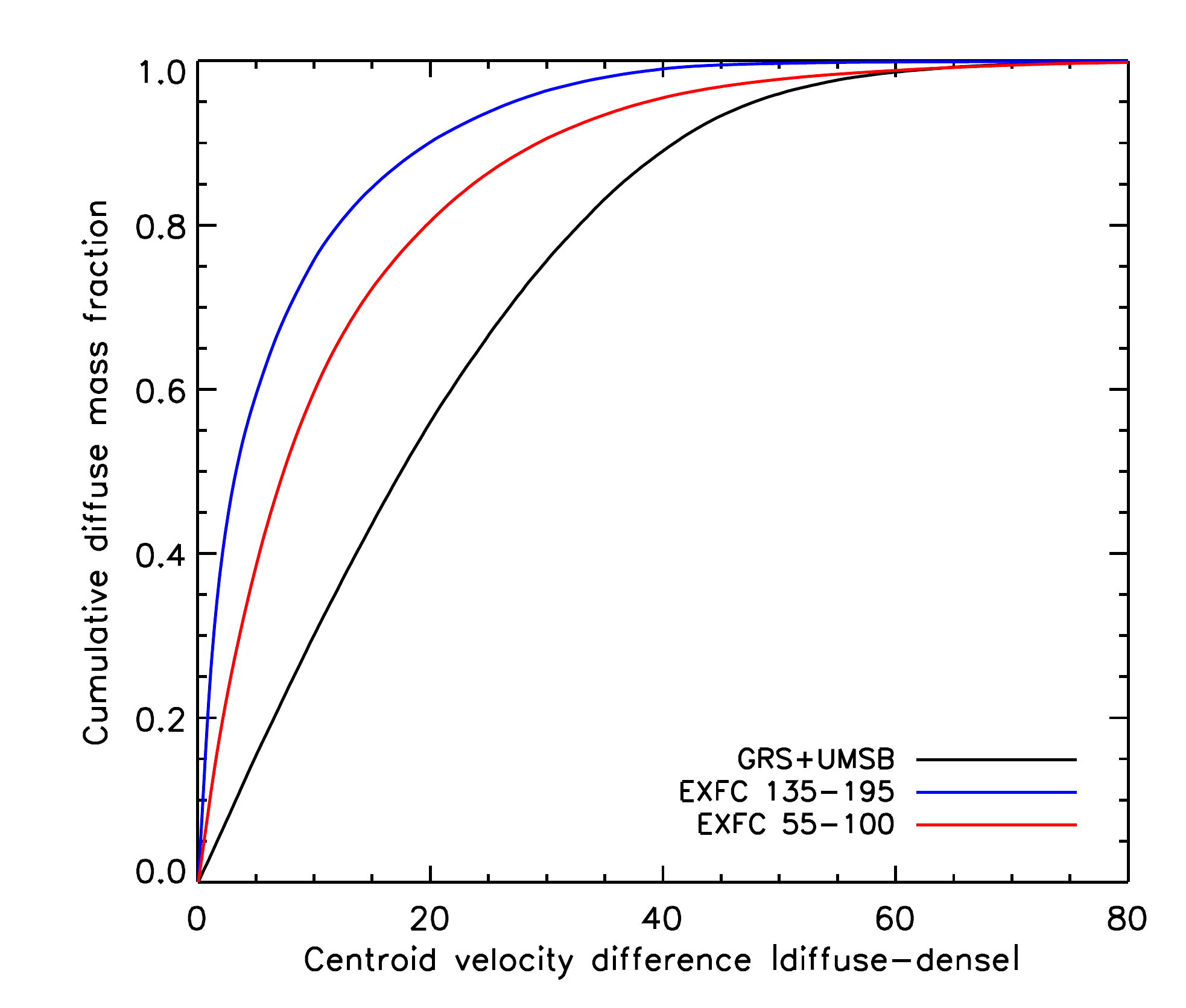} 

                          \caption{Cumulative mass fraction of gas classified as ``diffuse'' as a function of the centroid velocity difference between the ``diffuse'' and ``dense'' gas, calculated in sight-lines where both ``diffuse'' and ``dense'' gas components are detected. The black, red, and blue curves correspond to the GRS+UMSB, EXFC 55-100, and EXFC 135-195, respectively.}
\label{centroid_distrib}
\end{figure}

\indent We identify the ``diffuse'' gas reported here effectively based on its high $T_{12}/T_{13}$ ratio, which implies a low optical depth, and therefore a low surface density. In Section \ref{det_noise_section}, we estimate that the spectral surface density transition between the gas components we classify as ``dense'' and ``diffuse'' is about 10 \Msu pc$^{-2}$ (km s$^{-1}$)$^{-1}$, corresponding to surface densities of 25-50 \Msu pc$^{-2}$ for typical line widths. In this context, we interpret the ``diffuse'' gas as being of low surface density, and likely gravitationally unbound and unable to form stars, while the ``dense'' gas corresponds to a gas component the physical properties (density, surface density, viral parameter) of which are similar to those in the classical sense of molecular clouds. The ``diffuse'' gas is observed both in the form of isolated extended structures, but also in the envelopes of dense gas. \\
\indent There are, however, other effects that can induce high $T_{12}/T_{13}$ ratios. In particular, the wings of optically thick \COTs emission from dense clouds can be broader than the corresponding \COs line while emanating from the same dense gas.  In this study, the emission corresponding to those optically thick \COTs line wings would be included in the ``diffuse'' component. Thus, we may be overestimating the emission and mass of truly diffuse gas, while underestimating the amount of truly dense gas. We cannot differentiate the emission from truly diffuse gas from the dense gas emission in the opacity-broadened wings of the \COTs line, because we do not segment the emission into clouds (there are no GMCs in our study). However, we observe that 40\% of the gas mass classified here as ``diffuse'' in the outer Galaxy is located in sight-lines toward which no dense gas is detected. This implies that at least 40\% of the ``diffuse'' gas mass fraction reported here in the outer Galaxy corresponds to truly diffuse gas. In the inner Milky Way, the ``diffuse'' gas mass fraction with no associated dense component is 15\% in the EXFC 55-100 coverage, and 5\% in the GRS+UMSB coverage. However, these numbers are not meaningful in the inner Milky Way because most line-of-sights exhibit more than one CO line detection.\\
\indent To evaluate more quantitatively the fraction of gas that we report to be ``diffuse'', but actually corresponds to the opacity-broadened line wings of dense gas in sight-lines where both diffuse and dense gas are detected, we compute the centroid velocity maps of our ``diffuse'' and ``dense'' components. For sight-lines in which both ``diffuse'' and ``dense'' gas are detected, we then compute, for each survey, the cumulative mass distribution of ``diffuse'' gas as a function of the difference in centroid velocity between that ``diffuse'' and the  ``dense'' gas along the same line-of-sight (Figure \ref{centroid_distrib}). If the high $T_{12}/T_{13}$ ratio gas that we classify as ``diffuse'' actually corresponds to the optically thick line-wings of \COT, then one would expect the centroid velocity of this component to be similar to the centroid velocity of the ``dense'' gas. In the GRS+UMSB surveys, 90\% of the gas we report as ``diffuse'' has a centroid velocity farther than 5 \kms from the centroid of the ``dense'' gas along the same sight-line. In the EXFC 55-100 and EXFC 135-195, 65\% and 45\% of the gas mass that we classify as ``diffuse'' has a centroid velocity farther than 5 \kms from the centroid of the ``dense'' gas. The typical line width of CO clouds is 5 \kmsn, and so Figure \ref{centroid_distrib} implies that most of the gas mass classified as ``diffuse'' in this study does correspond to truly diffuse gas, and not to emission from the opacity-broadened line-wings of the \COTs line.

\indent  Because of the high critical density of CS emission and because we detect CS 2-1 emission with $T_{12}/T_{\mathrm{CS}}$ $<$ 15, corresponding to spectral surface densities $>$ 20 \Msu pc$^{-2}$ (km s$^{-1}$)$^{-1}$ and surface densities $>$ 100 \Msu pc$^{-2}$ for typical line-widths, we interpret the ``very dense'' gas as being relatively compact, gravitationally bound and star-forming  gas. Indeed, \citet{lada10} derive a threshold of 120 \Msu pc$^{-2}$ for star-formation to occur, close to the ``dense''-``very dense'' threshold used here. It is however worth mentioning that Galactic Plane surveys of the CS line with higher sensitivity than the GRS \citep{liszt95, Helfer97} have shown that every \COs feature has emission from CS at a level $\sim$ 1-2\% of the \COTs emission. The densities derived from such weak emission are consistent with rather diffuse molecular gas (low hundreds cm$^{-3}$). This is however not the gas we are probing here with CS emission at the level $T_{12}/T_{\mathrm{CS}}$ $<$ 15.

\subsection{Variable CO-H$_2$ conversion factor}

\indent In this work, we compute H$_2$ masses and surface densities in voxels with \COTs and \COs emission detectable at $>$2$\sigma$ under the assumption of LTE and a constant \COT/\COs abundance. We derive the (constant) conversion factor between \COTs luminosity and H$_2$ mass in this sample of voxels. For voxels with \COs emission below $2\sigma$ (and detected \COTs emission), we assume this same constant CO-to-H$_2$ conversion factor derived in the voxels detected in \COs (see Section \ref{properties_subsection}) to compute an H$_2$ mass. However, we note that the X$_{\mathrm{CO}}$ factor is likely to vary and increase for low-surface density gas. This gas is less shielded against the interstellar radiation field (ISRF), which affects CO more strongly than H$_2$. While H$_2$ is well protected against photodissociation above an extinction of $A_V \sim 1$, CO requires values of $A_V \approx 2 - 3$ under solar neighborhood conditions \citep[see, e.g.][]{tielens85, wolfire93, rollig07, glover10, glover11, shetty11a, shetty11b}. Indeed observations of nearby clouds show strong spatial variations of X$_{\mathrm{CO}}$ \citep[e.g., see][for a detailed analysis of the Perseus cloud]{lee14, pineda08}. The fraction of diffuse gas scales linearly with the CO-to-H$_2$ conversion factor assumed. Since the CO-to-H$_2$ conversion factor may be significantly higher in diffuse gas compared to dense gas, the diffuse gas fraction derived here represents a lower limit. We note however that, when averaged over a large enough volume of the ISM or when focusing on the bulk of the molecular mass traced by CO lines, taking a roughly constant X$_{\mathrm{ CO}}$-factor gives acceptable results for solar-metallicity galaxies even when applied to the diffuse component \citep{solomon87, young91, liszt10}.

\subsection{Diffuse CO gas and star-formation}

\indent As alluded to in Section \ref{introduction}, there is a debate about the universality and slope of the KS relation in nearby galaxies. \citet{shetty13, shetty14a} argue in favor of galaxy to galaxy variation. Most, but not all, galaxies in their study portray a sub-linear relation between star formation rate surface density and molecular gas surface density \citep[see also,][]{blanc09, ford13}. \citet{shetty14b} suggest that a non-linear KS relation may result from the presence of CO not related to dense star-forming clouds, perhaps in a diffuse but pervasive molecular component. Our analysis confirms that this component of the ISM exists and contains about 25\% of the total molecular ISM as traced by CO. If the SFR is linearly related to the amount of very high density gas, and if the radial trends we find in this work hold throughout the Galaxy, according to \citet{shetty14b}, the underlying relationship between the star formation rate surface density and $H_2$ would be super-linear (see their Fig. 3). We furthermore note that even in the inner Galaxy, which is clearly dominated by dense molecular gas, only 14\% of molecular gas is associated with known molecular clouds as identified in the UMSB+GRS surveys \citep{romanduval2010}. The bulk of this dense gas is found in a more distributed configuration. Our analysis suggests that the star formation process could simply be limited by the availability of such high density gas at any given time.

\section{Conclusion}\label{conclusion_section}

\indent We have examined the spatial distribution of three CO-emitting gas components in the Milky Way, a diffuse component traced by \COT, but dark in \CO, a dense component traced by both \COTs and \CO, and in the inner Galaxy only, a very dense component bright in \COT, \CO, and CS rotational emission. We have developed a robust algorithm to determine whether a voxel has significant emission from those tracers. The algorithm first smoothes the spectral cubes so that the S/N of the different line tracers are consistent with each other. A mask is then based on the thresholding (1$\sigma$) of the smoothed cubes. The detection masks are eroded and dilated to remove spurious noise peaks, since the 1$\sigma$ threshold only filters out 84\% of the noise. Finally, we apply the masks to the original (un-smoothed) spectral cubes. We have demonstrated that our approach accurately identifies all the low-level CO and CS emission. \\
\indent We have applied this detection algorithm to \COT, \CO, and CS spectral observations of the Milky Way in the GRS, UMSB, and EXFC surveys, and identified voxels with noise, diffuse, dense, or very dense CO emission. With kinematic distances to each voxel in the survey, we have derived masses and luminosities at every position in the Galaxy for each CO component. This allowed us to derive total masses of $1.5 \times 10^8$ \Msun, 4.9$\times 10^8$, and 2.9$\times 10^7$ \Msu for the diffuse, dense, and very dense components, respectively. Altogether, the diffuse gas comprises 25\% of the total molecular gas mass. The very dense gas represents 14\% of the total molecular gas mass. \\
\indent We have also derived the radial mass distributions of the three CO components. The surface density of molecular gas decreases by two orders of magnitude between Galactocentric radii of 3 kpc and 15 kpc. The dense CO gas dominates in mass in the inner Galaxy, with a dense gas fraction ranging from 90\% at $R_{gal}$ $=$ 4 kpc down to 50\% at the solar circle. The diffuse and dense gas has similar relative contributions in the outer Galaxy. The very dense gas fraction in the inner Galaxy appears to vary considerably with position. Locally in density peaks, the very dense gas fraction can reach 50\%. But the spatial distribution of the very dense gas is sparse, rendering its global mass contribution very small. Both the dense and very dense gas mass fractions are positively correlated with surface density. The overall radial distribution of CO gas in the Milky Way is consistent with previous studies based on coarser surveys summarized in the review article by \citet{heyer15}.\\
\indent We have derived the vertical distribution of molecular gas in the Milky Way as a function of galactocentric radius. In the inner Milky Way, the vertical molecular profiles are nearly Gaussian and dominated by the dense gas, with a FWHM of 110 pc. The very dense gas is much more concentrated on the Galactic plane, with a FWHM of $\sim$ 50 pc. In the outer Galaxy, the vertical molecular profiles are complex and multi-peaked, and wider than in the inner Milky Way, with FWHM as high as 300 pc. The vertical distribution and warp of CO molecular gas are also consistent with previous studies summarized in \citet{heyer15}.

\bibliographystyle{/users/duval/stsci_research/bibtex/apj11}
\bibliography{/Users/duval/work_bu/bibliography_nov_2010}

\begin{thebibliography}{}
\expandafter\ifx\csname natexlab\endcsname\relax\def\natexlab#1{#1}\fi

\bibitem[{{Battisti} \& {Heyer}(2014)}]{battisti14}
{Battisti}, A.~J., \& {Heyer}, M.~H. 2014, \apj, 780, 173

\bibitem[{{Bigiel} {et~al.}(2008){Bigiel}, {Leroy}, {Walter}, {Brinks}, {de
  Blok}, {Madore}, \& {Thornley}}]{bigiel08}
{Bigiel}, F., {Leroy}, A., {Walter}, F., {et~al.} 2008, \aj, 136, 2846

\bibitem[{{Blanc} {et~al.}(2009){Blanc}, {Heiderman}, {Gebhardt}, {Evans}, \&
  {Adams}}]{blanc09}
{Blanc}, G.~A., {Heiderman}, A., {Gebhardt}, K., {Evans}, II, N.~J., \&
  {Adams}, J. 2009, \apj, 704, 842

\bibitem[{{Burton} {et~al.}(1978){Burton}, {Liszt}, \& {Baker}}]{burton78}
{Burton}, W.~B., {Liszt}, H.~S., \& {Baker}, P.~L. 1978, \apjl, 219, L67

\bibitem[{{Clemens}(1985)}]{clemens85}
{Clemens}, D.~P. 1985, \apj, 295, 422

\bibitem[{{Clemens} {et~al.}(1986){Clemens}, {Sanders}, {Scoville}, \&
  {Solomon}}]{clemens86}
{Clemens}, D.~P., {Sanders}, D.~B., {Scoville}, N.~Z., \& {Solomon}, P.~M.
  1986, \apjs, 60, 297

\bibitem[{{Dame} \& {Thaddeus}(1994)}]{dame94}
{Dame}, T.~M., \& {Thaddeus}, P. 1994, \apjl, 436, L173

\bibitem[{{Ford} {et~al.}(2013){Ford}, {Gear}, {Smith}, {Eales}, {Baes},
  {Bendo}, {Boquien}, {Boselli}, {Cooray}, {De Looze}, {Fritz}, {Gentile},
  {Gomez}, {Gordon}, {Kirk}, {Lebouteiller}, {O'Halloran}, {Spinoglio},
  {Verstappen}, \& {Wilson}}]{ford13}
{Ford}, G.~P., {Gear}, W.~K., {Smith}, M.~W.~L., {et~al.} 2013, \apj, 769, 55

\bibitem[{{Glover} {et~al.}(2010){Glover}, {Federrath}, {Mac Low}, \&
  {Klessen}}]{glover10}
{Glover}, S.~C.~O., {Federrath}, C., {Mac Low}, M., \& {Klessen}, R.~S. 2010,
  \mnras, 404, 2

\bibitem[{{Glover} \& {Mac Low}(2011)}]{glover11}
{Glover}, S.~C.~O., \& {Mac Low}, M.-M. 2011, \mnras, 412, 337

\bibitem[{{Goldsmith} {et~al.}(2008){Goldsmith}, {Heyer}, {Narayanan}, {Snell},
  {Li}, \& {Brunt}}]{goldsmith08}
{Goldsmith}, P.~F., {Heyer}, M., {Narayanan}, G., {et~al.} 2008, \apj, 680, 428

\bibitem[{{Heiderman} {et~al.}(2010){Heiderman}, {Evans}, {Allen}, {Huard}, \&
  {Heyer}}]{heiderman10}
{Heiderman}, A., {Evans}, II, N.~J., {Allen}, L.~E., {Huard}, T., \& {Heyer},
  M. 2010, \apj, 723, 1019

\bibitem[{{Helfer} \& {Blitz}(1997)}]{Helfer97}
{Helfer}, T.~T., \& {Blitz}, L. 1997, \apj, 478, 233

\bibitem[{{Heyer} \& {Dame}(2015)}]{heyer15}
{Heyer}, M., \& {Dame}, T.~M. 2015, \araa, 53, 583

\bibitem[{{Heyer} {et~al.}(2009){Heyer}, {Krawczyk}, {Duval}, \&
  {Jackson}}]{heyer09}
{Heyer}, M., {Krawczyk}, C., {Duval}, J., \& {Jackson}, J.~M. 2009, \apj, 699,
  1092

\bibitem[{{Jackson} {et~al.}(2002){Jackson}, {Bania}, {Simon}, {Kolpak},
  {Clemens}, \& {Heyer}}]{jackson02}
{Jackson}, J.~M., {Bania}, T.~M., {Simon}, R., {et~al.} 2002, \apjl, 566, L81

\bibitem[{{Kennicutt}(1998)}]{kennicutt98}
{Kennicutt}, Jr., R.~C. 1998, \apj, 498, 541

\bibitem[{{Klessen} \& {Glover}(2014)}]{Klessen14}
{Klessen}, R.~S., \& {Glover}, S.~C.~O. 2014, ArXiv e-prints, arXiv:1412.5182

\bibitem[{{Knapp}(1974)}]{knapp74}
{Knapp}, G.~R. 1974, \aj, 79, 527

\bibitem[{{Krumholz} {et~al.}(2012){Krumholz}, {Dekel}, \&
  {McKee}}]{krumholz12}
{Krumholz}, M.~R., {Dekel}, A., \& {McKee}, C.~F. 2012, \apj, 745, 69

\bibitem[{{Lada} {et~al.}(2010){Lada}, {Lombardi}, \& {Alves}}]{lada10}
{Lada}, C.~J., {Lombardi}, M., \& {Alves}, J.~F. 2010, \apj, 724, 687

\bibitem[{{Lee} {et~al.}(2014){Lee}, {Stanimirovi{\'c}}, {Wolfire}, {Shetty},
  {Glover}, {Molina}, \& {Klessen}}]{lee14}
{Lee}, M.-Y., {Stanimirovi{\'c}}, S., {Wolfire}, M.~G., {et~al.} 2014, \apj,
  784, 80

\bibitem[{{Leroy} {et~al.}(2013){Leroy}, {Walter}, {Sandstrom}, {Schruba},
  {Munoz-Mateos}, {Bigiel}, {Bolatto}, {Brinks}, {de Blok}, {Meidt}, {Rix},
  {Rosolowsky}, {Schinnerer}, {Schuster}, \& {Usero}}]{leroy13}
{Leroy}, A.~K., {Walter}, F., {Sandstrom}, K., {et~al.} 2013, \aj, 146, 19

\bibitem[{{Liszt}(1995)}]{liszt95}
{Liszt}, H.~S. 1995, \apj, 442, 163

\bibitem[{{Liszt} {et~al.}(1984){Liszt}, {Burton}, \& {Xiang}}]{liszt84}
{Liszt}, H.~S., {Burton}, W.~B., \& {Xiang}, D.-L. 1984, \aap, 140, 303

\bibitem[{{Liszt} {et~al.}(2010){Liszt}, {Pety}, \& {Lucas}}]{liszt10}
{Liszt}, H.~S., {Pety}, J., \& {Lucas}, R. 2010, \aap, 518, A45

\bibitem[{{Liu} {et~al.}(2011){Liu}, {Koda}, {Calzetti}, {Fukuhara}, \&
  {Momose}}]{liu11}
{Liu}, G., {Koda}, J., {Calzetti}, D., {Fukuhara}, M., \& {Momose}, R. 2011,
  \apj, 735, 63

\bibitem[{{Mac Low} \& {Klessen}(2004)}]{maclow04}
{Mac Low}, M.-M., \& {Klessen}, R.~S. 2004, Reviews of Modern Physics, 76, 125

\bibitem[{{Milam} {et~al.}(2005){Milam}, {Savage}, {Brewster}, {Ziurys}, \&
  {Wyckoff}}]{milam05}
{Milam}, S.~N., {Savage}, C., {Brewster}, M.~A., {Ziurys}, L.~M., \& {Wyckoff},
  S. 2005, \apj, 634, 1126

\bibitem[{{Momose} {et~al.}(2013){Momose}, {Koda}, {Kennicutt}, {Egusa},
  {Calzetti}, {Liu}, {Donovan Meyer}, {Okumura}, {Scoville}, {Sawada}, \&
  {Kuno}}]{momose13}
{Momose}, R., {Koda}, J., {Kennicutt}, Jr., R.~C., {et~al.} 2013, \apjl, 772,
  L13

\bibitem[{{Neufeld} {et~al.}(2015){Neufeld}, {Godard}, {Gerin}, {Pineau des
  For{\^e}ts}, {Bernier}, {Falgarone}, {Graf}, {G{\"u}sten}, {Herbst},
  {Lesaffre}, {Schilke}, {Sonnentrucker}, \& {Wiesemeyer}}]{neufeld15}
{Neufeld}, D.~A., {Godard}, B., {Gerin}, M., {et~al.} 2015, \aap, 577, A49

\bibitem[{{Pety} {et~al.}(2013){Pety}, {Schinnerer}, {Leroy}, {Hughes},
  {Meidt}, {Colombo}, {Dumas}, {Garc{\'{\i}}a-Burillo}, {Schuster}, {Kramer},
  {Dobbs}, \& {Thompson}}]{pety13}
{Pety}, J., {Schinnerer}, E., {Leroy}, A.~K., {et~al.} 2013, \apj, 779, 43

\bibitem[{{Pineda} {et~al.}(2008){Pineda}, {Caselli}, \& {Goodman}}]{pineda08}
{Pineda}, J.~E., {Caselli}, P., \& {Goodman}, A.~A. 2008, \apj, 679, 481

\bibitem[{{Rathborne} {et~al.}(2009){Rathborne}, {Johnson}, {Jackson}, {Shah},
  \& {Simon}}]{rathborne09}
{Rathborne}, J.~M., {Johnson}, A.~M., {Jackson}, J.~M., {Shah}, R.~Y., \&
  {Simon}, R. 2009, \apjs, 182, 131

\bibitem[{{R{\"o}llig} {et~al.}(2007){R{\"o}llig}, {Abel}, {Bell}, {Bensch},
  {Black}, {Ferland}, {Jonkheid}, {Kamp}, {Kaufman}, {Le Bourlot}, {Le Petit},
  {Meijerink}, {Morata}, {Ossenkopf}, {Roueff}, {Shaw}, {Spaans}, {Sternberg},
  {Stutzki}, {Thi}, {van Dishoeck}, {van Hoof}, {Viti}, \&
  {Wolfire}}]{rollig07}
{R{\"o}llig}, M., {Abel}, N.~P., {Bell}, T., {et~al.} 2007, \aap, 467, 187

\bibitem[{{Roman-Duval} {et~al.}(2009){Roman-Duval}, {Jackson}, {Heyer},
  {Johnson}, {Rathborne}, {Shah}, \& {Simon}}]{romanduval2009}
{Roman-Duval}, J., {Jackson}, J.~M., {Heyer}, M., {et~al.} 2009, \apj, 699,
  1153

\bibitem[{{Roman-Duval} {et~al.}(2010){Roman-Duval}, {Jackson}, {Heyer},
  {Rathborne}, \& {Simon}}]{romanduval2010}
{Roman-Duval}, J., {Jackson}, J.~M., {Heyer}, M., {Rathborne}, J., \& {Simon},
  R. 2010, \apj, 723, 492

\bibitem[{{Rosolowsky} \& {Leroy}(2006)}]{rosolowsky06}
{Rosolowsky}, E., \& {Leroy}, A. 2006, \pasp, 118, 590

\bibitem[{{Rosolowsky} {et~al.}(2008){Rosolowsky}, {Pineda}, {Kauffmann}, \&
  {Goodman}}]{rosolowsky08}
{Rosolowsky}, E.~W., {Pineda}, J.~E., {Kauffmann}, J., \& {Goodman}, A.~A.
  2008, \apj, 679, 1338

\bibitem[{{Sanders} {et~al.}(1986){Sanders}, {Clemens}, {Scoville}, \&
  {Solomon}}]{sanders86}
{Sanders}, D.~B., {Clemens}, D.~P., {Scoville}, N.~Z., \& {Solomon}, P.~M.
  1986, \apjs, 60, 1

\bibitem[{{Schmidt}(1959)}]{schmidt59}
{Schmidt}, M. 1959, \apj, 129, 243

\bibitem[{{Shetty} {et~al.}(2014{\natexlab{a}}){Shetty}, {Clark}, \&
  {Klessen}}]{shetty14b}
{Shetty}, R., {Clark}, P.~C., \& {Klessen}, R.~S. 2014{\natexlab{a}}, \mnras,
  442, 2208

\bibitem[{{Shetty} {et~al.}(2011{\natexlab{a}}){Shetty}, {Glover}, {Dullemond},
  \& {Klessen}}]{shetty11a}
{Shetty}, R., {Glover}, S.~C., {Dullemond}, C.~P., \& {Klessen}, R.~S.
  2011{\natexlab{a}}, \mnras, 412, 1686

\bibitem[{{Shetty} {et~al.}(2011{\natexlab{b}}){Shetty}, {Glover}, {Dullemond},
  {Ostriker}, {Harris}, \& {Klessen}}]{shetty11b}
{Shetty}, R., {Glover}, S.~C., {Dullemond}, C.~P., {et~al.} 2011{\natexlab{b}},
  \mnras, 415, 3253

\bibitem[{{Shetty} {et~al.}(2013){Shetty}, {Kelly}, \& {Bigiel}}]{shetty13}
{Shetty}, R., {Kelly}, B.~C., \& {Bigiel}, F. 2013, \mnras, 430, 288

\bibitem[{{Shetty} {et~al.}(2014{\natexlab{b}}){Shetty}, {Kelly}, {Rahman},
  {Bigiel}, {Bolatto}, {Clark}, {Klessen}, \& {Konstandin}}]{shetty14a}
{Shetty}, R., {Kelly}, B.~C., {Rahman}, N., {et~al.} 2014{\natexlab{b}},
  \mnras, 437, L61

\bibitem[{{Solomon} \& {Rivolo}(1989)}]{solomon89}
{Solomon}, P.~M., \& {Rivolo}, A.~R. 1989, \apj, 339, 919

\bibitem[{{Solomon} {et~al.}(1987){Solomon}, {Rivolo}, {Barrett}, \&
  {Yahil}}]{solomon87}
{Solomon}, P.~M., {Rivolo}, A.~R., {Barrett}, J., \& {Yahil}, A. 1987, \apj,
  319, 730

\bibitem[{{Stutzki}(2014)}]{stutzki14}
{Stutzki}, J. 2014, {GAUSSCLUMPS: Gaussian-shaped clumping from a spectral
  map}, Astrophysics Source Code Library, ascl:1406.018

\bibitem[{{Tielens} \& {Hollenbach}(1985)}]{tielens85}
{Tielens}, A.~G.~G.~M., \& {Hollenbach}, D. 1985, \apj, 291, 722

\bibitem[{{Williams} {et~al.}(1994){Williams}, {de Geus}, \&
  {Blitz}}]{williams94}
{Williams}, J.~P., {de Geus}, E.~J., \& {Blitz}, L. 1994, \apj, 428, 693

\bibitem[{{Williams} \& {McKee}(1997)}]{williams97}
{Williams}, J.~P., \& {McKee}, C.~F. 1997, \apj, 476, 166

\bibitem[{{Wolfire} {et~al.}(1993){Wolfire}, {Hollenbach}, \&
  {Tielens}}]{wolfire93}
{Wolfire}, M.~G., {Hollenbach}, D., \& {Tielens}, A.~G.~G.~M. 1993, \apj, 402,
  195

\bibitem[{{Young} \& {Scoville}(1991)}]{young91}
{Young}, J.~S., \& {Scoville}, N.~Z. 1991, \araa, 29, 581

\end{thebibliography}

\acknowledgments{RS and RSK acknowledge support from the Deutsche Forschungsgemeinschaft (DFG) for funding through the SPP 1573 ``The Physics of the Interstellar Medium'' as well as via SFB 881 ``The Milky Way System'' (sub-projects B1, B2 and B8). RSK also receives funding from the European Research Council under the European CommunityÕs Seventh Framework Program (FP7/2007-2013) via the ERC Advanced Grant ``STARLIGHT'' (project number 339177). 

}

\end{document}